\documentclass[11pt]{article}

\usepackage{lineno}
\usepackage{hyperref}
\usepackage{amssymb}
\usepackage{amsmath}
\usepackage{mathtools}

\modulolinenumbers[5]

\begin{document}

\newtheorem{theorem}{Theorem}[section]
\newtheorem{lemma}[theorem]{Lemma}
\newtheorem{corollary}[theorem]{Corollary}
\newtheorem{proposition}[theorem]{Proposition}
\newtheorem{observation}[theorem]{Observation}
\newcommand{\blackslug}{\penalty 1000\hbox{
    \vrule height 8pt width .4pt\hskip -.4pt
    \vbox{\hrule width 8pt height .4pt\vskip -.4pt
          \vskip 8pt
      \vskip -.4pt\hrule width 8pt height .4pt}
    \hskip -3.9pt
    \vrule height 8pt width .4pt}}
\newcommand{\proofend}{\quad\blackslug}
\newenvironment{proof}{$\;$\newline \noindent {Proof.}$\;$\rm}{\qad}
\newcommand{\qad}{\hspace*{\fill}\blackslug}
\newenvironment{definition}{$\;$\newline \noindent {\bf Definition}$\;$}{$\;$\newline}
\def\boxit#1{\vbox{\hrule\hbox{\vrule\kern4pt
  \vbox{\kern1pt#1\kern1pt}
\kern2pt\vrule}\hrule}}

\newcommand{\rat}{\mathbb{Q}}
\newcommand{\real}{\mathbb{R}}
\newcommand{\nat}{\mathbb{N}}
\newcommand{\umatch}{p-{\sc Matching}}
\newcommand{\wmatch}{p-{\sc Generalized-WT-Matching}}
\newcommand{\wmatcheq}{p-{\sc wMatching}}
\newcommand{\SSS}{\mathcal{S}}
\newcommand{\TT}{{\mathcal Trim}}
\newcommand{\RR}{{\mathcal Reduce}}
\newcommand{\CC}{{\mathcal C}}
\newcommand{\RC}{{\mathcal R}}
\newcommand{\NP}{\mbox{{\sf NP}}}

\newcommand{\plog}{\mbox{\rm polylog}(n)}


\title{
{\bf Streaming Algorithms for Graph $k$-Matching\\with Optimal or Near-Optimal Update Time}}

\author{{\sc Jianer Chen}\footnote{Department of Computer Science and Engineering, 
    Texas A\&M University, College Station, TX 77843, USA (chen@cse.tamu.edu, huangqin@tamu.edu).} , \ 
  {\sc Qin Huang}\footnotemark[1] , \  
  {\sc Iyad Kanj}\footnote{School of Computing, DePaul University, Chicago, 
     IL 60604, USA (ikanj@depaul.edu).} , \   
   {\sc Qian Li}\footnote{Institute of Computing Technology, Chinese Academy of Sciences, Beijing, China
     (liqian@ict.ac.cn).} , \  {\sc and} 
   {\sc Ge Xia}\footnote{Department of Computer Science, Lafayette College, Easton, PA 18042, USA 
     (xiag@lafayette.edu).}
}

\date{}

\maketitle

\begin{abstract} 
We present streaming algorithms for the graph $k$-matching problem in both the insert-only and dynamic 
models. Our algorithms, with space complexity matching the best upper bounds, have optimal or near-optimal 
update time, significantly improving on previous results. More specifically, for the insert-only streaming model, 
we present a one-pass algorithm with optimal space complexity $O(k^2)$ and optimal update time $O(1)$, 
that \emph{w.h.p.}~(with high probability) computes a maximum weighted 
$k$-matching of a given weighted graph. The update time of our algorithm significantly improves the 
previous upper bound of $O(\log k)$, which was derived only for $k$-matching on unweighted graphs. 
For the dynamic streaming model, we present a one-pass algorithm that \emph{w.h.p.}~computes a 
maximum weighted $k$-matching in $O(Wk^2 \cdot \plog)$ space\footnote{We denote by $\plog$ the 
function $\log^{O(1)} n$, where $n$ is the size of the input.} and with $O(\plog)$ update time, where $W$ 
is the number of distinct edge weights. Again the update time of our algorithm improves the previous 
upper bound of $O(k^2 \cdot \plog)$. This algorithm, when applied to unweighted graphs, gives a 
streaming algorithm on the dynamic model whose space and update time complexities are 
both near-optimal. Our results also imply a streaming approximation algorithm for maximum weighted 
$k$-matching whose space complexity matches the best known upper bound with a significantly improved 
update time. 

\vspace{2mm}

\noindent {\bf keywords.}\  streaming algorithm; graph matching; parameterized algorithm; lower bound
\end{abstract} 


\section{Introduction}
\label{subsec:def}

Streaming algorithms for graph matching have been studied extensively. A graph \emph{stream} $\SSS$ 
for an underlying graph $G$ is a sequence of edge operations. In the \emph{insert-only} streaming 
model, each operation is an edge-insertion, while in the \emph{dynamic} streaming model each operation 
is either an edge-insertion or an edge-deletion (with a specified weight if $G$ is weighted). Most of the 
previous work on the graph matching problem in the streaming model have focused on approximating a 
maximum matching, with the majority of the work pertaining to the (simpler) insert-only  model (see, e.g., 
\cite{soda17,soda16-31,soda16-32,soda-21,mcgregor,new-soda-17}). More 
recently, streaming algorithms for the {\sc Graph $k$-Matching} problem (i.e., constructing a matching of 
$k$ edges in an unweighted graph or a maximum weighted matching of $k$ edges in a weighted graph), 
in both the insert-only and the dynamic models, have drawn increasing interests 
\cite{bury,Chitnis2016, spaa2015, soda15, soda16-22}. 

The performance of streaming algorithms is measured by the limited memory ({\it space}) and the limited 
processing time per item ({\it update time}). For the space complexity, a lower bound of $\Omega(k^2)$ has 
been known for {\sc Graph $k$-Matching} on unweighted graphs for randomized streaming algorithms, 
even in the simpler insert-only model~\cite{Chitnis2016}. Nearly space-optimal streaming algorithms, i.e., 
streaming algorithms with space complexity $O(k^2 \cdot \plog)$, have been developed for {\sc Graph 
$k$-Matching} on unweighted graphs~\cite{bury,Chitnis2016}. 

The current paper is focused on the update time of streaming algorithms for {\sc Graph $k$-Matching}. 
While there has been much work pertaining to the space complexity of streaming algorithms for graph 
matching, much less is known about the update time complexity of the problem. Note that the update time 
sometimes could be even more important than the space complexity~\cite{lnn15}, since the data stream 
can come at a very high rate. If the update processing rate does not catch the update arrival rate, the 
whole system may fail (see, e.g., \cite{ay20,tz12}). A major contribution of the current paper is the 
development of a collection of streaming algorithms for {\sc Graph $k$-Matching} that, while keeping 
the optimal or near-optimal space complexity, also reach optimal or near-optimal update time. 

\subsection{\bf Previous work on Graph $k$-Matching}

We start by reviewing the relevant previous work on the problem.  

Fafianie and Kratsch~\cite{soda16-22} studied kernelization streaming algorithms in the insert-only model 
for the \NP-hard $d$-{\sc Set Matching} problem (among others), which for $d=2$, is equivalent to the 
{\sc Graph $k$-Matching} problem on unweighted graphs. Their result gives a one-pass (deterministic) 
kernelization streaming algorithm for {\sc Graph $k$-Matching} on unweighted graphs. The algorithm 
implies a streaming algorithm in the insert-only model for the {\sc Graph $k$-Matching} problem on 
unweighted graphs, with space complexity $O(k^2)$ and update time $O(\log k)$.

More recently, streaming algorithms in the dynamic model for the {\sc Graph $k$-Matching} problem have 
been studied \cite{bury,Chitnis2016,spaa2015,soda15}. Under the assumption that at \emph{every} instant 
the size (i.e., the cardinality) of a maximum matching of the graph stream is bounded by $k$, a randomized 
one-pass dynamic streaming algorithm is given in~\cite{soda15}, which was refined in \cite{spaa2015}. The 
algorithm \emph{w.h.p.} computes a maximum matching in an unweighted graph stream, and runs in 
$O(k^2 \cdot \plog)$ space and $O(k^2 \cdot \plog)$ update time (see \cite{soda15}).

The authors of~\cite{Chitnis2016} revisited the problem of constructing maximum matchings in the 
dynamic streaming model. Under a slightly less restricted constraint that the size of a maximum 
matching of the stream graph is bounded by $k$ (we will call this constraint ``the Size-$k$ Constraint''), 
a sketch-based streaming algorithm is presented in \cite{Chitnis2016} that \emph{w.h.p.}~computes a 
maximum matching of an unweighted graph. The algorithm retains the space complexity at $O(k^2 \cdot \plog)$ 
but has an improved update time of $O(\plog)$.  

For general graph streams that may not satisfy the Size-$k$ Constraint, a randomized approximation 
algorithm was given in~\cite{Chitnis2016} for maximum matchings in unweighted graph streams. Specifically, 
if the graph contains matchings of size larger than $k$, then for any $1 \le \alpha \le \sqrt{k}$ and 
$0<\epsilon \le 1$, there exists an $O(k^2/(\alpha^3\epsilon^2) \cdot \plog)$-space algorithm that 
\emph{w.h.p.} returns a matching of size at least $(1-\epsilon)k/(2\alpha)$. The algorithm has 
$O(k^2 \cdot \plog/(\alpha^2\epsilon^2))$ update time (see \cite{Chitnis2016}, Theorem 4.1). In particular, 
for a graph $G$ in the stream in which the size of a maximum matching is at least $c_0k$ for a constant 
$c_0 \geq 1$, the algorithm, by properly choosing $\alpha$ and $\epsilon$ (e.g., if 
$c_0 = 4$ then let $\alpha = 1$ and $\epsilon = 1/2$), will \emph{w.h.p.}~construct a matching of size 
at least $k$ in the graph $G$. This streaming algorithm has space complexity $O(k^2 \cdot \plog)$ but its 
update time is raised back to $O(k^2 \cdot \plog)$. When we combine this algorithm with the streaming 
algorithm under the Size-$k$ Constraint (more precisely, under the Size-$(c_0 k)$ Constraint) given in 
\cite{Chitnis2016}, we will obtain a streaming algorithm in the dynamic model for {\sc Graph $k$-Matching} 
on general unweighted graph streams (i.e., without the assumption of the Size-$k$ Constraint), which 
runs in space complexity $O(k^2 \cdot \plog)$ and update time $O(k^2 \cdot \plog)$. Since $\Omega(k^2)$ is 
a lower bound on the space complexity of streaming algorithms for the {\sc Graph $k$-Matching} problem 
\cite{soda15}, the streaming algorithm described above for {\sc Graph $k$-Matching} has near-optimal 
space complexity (i.e., optimal modulo a poly-logarithmic factor). 

As described in \cite{Chitnis2016}, streaming algorithms for {\sc Graph $k$-Matching} on unweighted 
graph streams can be extended to solve the {\sc Graph $k$-Matching} problem on weighted graph 
streams (i.e., constructing a maximum weighted matching of $k$ edges in a weighted graph stream), 
with space complexity increased by a factor of the number $W$ of distinct edge weights. Thus, under 
the Size-$k$ Constraint, there is a streaming algorithm in the dynamic model for the {\sc Graph 
$k$-Matching} problem on weighted graph streams with space complexity $O(k^2 W \cdot \plog)$ and 
update time $O(\plog)$, while without the assumption of the Size-$k$ Constraint, there is a 
streaming algorithm in the dynamic model for the {\sc Graph $k$-Matching} problem on weighted graph 
streams with space complexity $O(k^2 W \cdot \plog)$ and update time $O(k^2 \cdot \plog)$.

The above described algorithms are the best known streaming algorithms for the {\sc Graph 
$k$-Matching} problem. 

\subsection{Our contributions}

We start by discussing our results for the insert-only model. We present a one-pass randomized streaming 
algorithm that constructs a maximum weighted $k$-matching in a weighted graph. Our algorithm runs 
in $O(k^2)$ space and has $O(1)$ update time, which both are optimal. Our algorithm relies on 
the critical observation that there is a ``compact'' subgraph of size $O(k^2)$ that contains a maximum 
weighted $k$-matching in the original graph. We show that using techniques of universal hashing, 
\emph{w.h.p.}, we can identify the compact subgraph effectively, and that using the technique of interleaving 
executions of multiple parts of the algorithm, we can efficiently update the compact subgraph when new 
edges are inserted while keeping the $O(1)$ update time.  

Compared to the previous best result by Fafianie and Kratsch~\cite{soda16-22}, who developed a 
(deterministic) kernelization streaming algorithm that implies a one-pass streaming algorithm in the 
insert-only model for {\sc Graph $k$-Matching} on unweighted graphs with space complexity $O(k^2)$ 
and update time $O(\log k)$, our algorithm is randomized, achieving the same (optimal) space 
complexity, but also has optimal update time $O(1)$. Most significantly, our streaming algorithm 
solves the {\sc Graph $k$-Matching} problem on weighted graphs, which is a much more difficult problem 
compared to the problem on unweighted graphs. 

We then study streaming algorithms for {\sc Graph $k$-Matching} in the dynamic model. We give a one-pass 
randomized streaming algorithm that, for a weighted graph $G$ containing a $k$-matching, \emph{w.h.p.}, 
constructs a maximum weighted $k$-matching of $G$. The algorithm runs in $O(k^2 W \cdot \plog)$ space 
and has $O(\plog)$ update time, where $W$ is the number of distinct edge weights in the graph. This result 
directly implies a one-pass randomized streaming algorithm for {\sc Graph $k$-Matching} on unweighted 
graphs, with near-optimal space complexity $O(k^2 \cdot \plog)$ and near-optimal update time $O(\plog)$. 
 
The faster update time of our streaming algorithm, while keeping the same (near-optimal) space complexity, 
is achieved based on a technique of randomized construction of a many-to-many mapping between a 
given large set $U$ and a small integral interval. Note that this approach is different from that of previous 
randomized streaming algorithms, which in general {\it partition} the set $U$ into disjoint subsets. Briefly 
speaking, for any (unknown) $k$-subset $S$ of the set $U$ we construct a small collection $H^+$ of 
$O(\log k)$ hash functions, each using $O(k \cdot \plog)$ space. The collection $H^+$ makes a 
many-to-many mapping between $U$ and an integral interval $I$ of size $O(k \cdot \plog)$. We show 
that \emph{w.h.p.}~there are $k$ integers in $I$ whose pre-images in $U$ are pairwise disjoint and each 
contains exactly one element in $S$. Compared to the popular approach for graph matching, which uses 
universal hash functions for streaming algorithms, our approach uses less space and achieves $O(\plog)$ 
update time. This technique combined with the $\ell_0$-sampling techniques ~\cite{l0-sampling,pods11} 
enables us to select a smaller subset of edges, from the vertex subsets of our construction, that 
\emph{w.h.p.}~contains the desired $k$-matching. From this smaller subset of edges, a maximum weighted 
$k$-matching can be extracted.


In comparison with the previous best results, Chitnis et al.~\cite{Chitnis2016}, under the Size-$k$ Constraint, 
developed a randomized streaming algorithm in the dynamic model for {\sc Graph $k$-Matching} on 
unweighted graphs, that has the same space complexity $O(k^2 \cdot \plog)$ and update time $O(\plog)$  
as our algorithm. However, our algorithm is not restricted to the Size-$k$ Constraint, which seems a rather 
strong assumption on graph streams. The previous best streaming algorithm for {\sc Graph $k$-Matching} 
on unweighted graphs without the assumption of the Size-$k$ Constraint, as given in \cite{Chitnis2016} and 
explained above, runs in space $O(k^2 \cdot \plog)$ and has update time $O(k^2 \cdot \plog)$. Compared to this 
algorithm, our algorithm has a much faster update time of $O(\plog)$. Similarly, compared with the best 
streaming algorithms in the dynamic model for {\sc Graph $k$-Matching} on weighted graphs as given 
in \cite{Chitnis2016}, our algorithm, while keeping the space complexity matching that in \cite{Chitnis2016}, 
is applicable to a much larger class of graphs (i.e., without assuming the Size-$k$ Constraint), and has 
significantly improved (and near-optimal) update time $O(\plog)$. 
 
A byproduct of our results is a one-pass streaming approximation algorithm that, for any $\epsilon > 0$, 
\emph{w.h.p.}~computes a $k$-matching in a weighted graph stream that is within a factor of $1-\epsilon$ 
from a maximum weighted $k$-matching. The algorithm runs in $O(k^2 \log R_{\it wt} \cdot \plog/\epsilon)$ space 
and has $O(\plog)$ update time, where $R_{\it wt}$ is the ratio of the maximum edge-weight to the 
minimum edge-weight in the graph. This result improves the update time complexity over the approximation 
result in~\cite{Chitnis2016}, which has the same space complexity but has update time $O(k^2 \cdot \plog)$.

%
 
We mention that Chen et al.~\cite{chen2020} studied algorithms for $k$-matching in unweighted and 
weighted graphs in the RAM model with limited computational resources. Clearly, the RAM model is very 
different from the streaming model. In order to translate their algorithm to the streaming model, it would 
require $\Omega(n k)$ space and multiple passes, where $n$ is the number of vertices. However, we 
mention that one of the steps of our algorithm in the insert-only model was inspired by an operation for 
constructing a reduced graph, which was introduced in~\cite{chen2020}. 

Finally, there has been work on computing matchings in special graph classes, and with respect to parameters 
other than the cardinality of the matching (see, e.g.~\cite{cormode,Niedermeier4,mcgregor,Niedermeier2}).

\section{Preliminaries}
\label{sec:prelim}

For a positive integer $i$, let $[i]$ denote the set of integers $\{1, 2, \ldots, i\}$, and let $[i]^-$ denote 
the set $\{0, 1, \ldots, i-1\}$. We write ``{\it u.a.r.}'' as an abbreviation for ``uniformly at random''. 

All graphs discussed in this paper are undirected and simple. We write $V(G)$ and $E(G)$ for the 
vertex set and edge set of a graph $G$, respectively, and write $[u, v]$ for an edge with the endpoints 
$u$ and $v$. The {\it size} of a graph $G$, denoted by $|G|$, is equal to the number of vertices plus 
the number of edges in the graph. A \emph{matching} $M \subseteq E(G)$ is a set of edges in which 
no two edges share a common endpoint. A matching $M$ is a \emph{$k$-matching} if it consists of 
exactly $k$ edges. A weighted graph $G$ is a graph associated with a weight function 
$wt: E(G) \longrightarrow \mathbb{R}$; we denote the weight of an edge $e$ by $wt(e)$. Let $M$ be a 
matching in a weighted graph $G$. The weight of $M$, $wt(M)$, is the sum of the weights of the edges 
in $M$.  A \emph{maximum weighted $k$-matching} in a weighted graph $G$ is a $k$-matching whose 
weight is the maximum over all $k$-matchings in $G$. 

\subsection{The graph streaming model} 
\label{subsec:graphstream}

A graph \emph{stream} $\SSS$ for an underlying graph $G$ is a sequence of {\it elements}, each of 
the form $(e, op)$, where $op$ is an {\it update} to an edge $e$ in $G$. An update could be either an 
\emph{insertion} or a \emph{deletion} of an edge (and would include the edge weight if the graph $G$ is 
weighted). In the \emph{insert-only graph streaming} model, a  graph $G$ is given as a stream 
$\SSS$ of elements in which each operation is an edge insertion, while in the \emph{dynamic graph 
streaming} model a graph $G$ is given as a stream $\SSS$ of elements in which the operations 
could be either edge insertions or edge deletions. We will assume that a graph stream always starts 
with an empty edge set. 

Without loss of generality, we will assume that a graph $G$ of $n$ vertices has $[n]^-$ as its vertex set, 
and that the length of a stream $\SSS$ for $G$ is polynomial in $n$. Since the  graph $G$ can have at 
most $n(n-1)/2$ edges, each edge in $G$ can be represented as a unique number in $[n(n-1)/2]^-$. 

\subsection{Problem definitions}
\label{subsec:problemdefinition}

The formulation of our problem is of a multivariate nature. An instance of our problem is of the form 
$(\SSS, k)$, where $\SSS$ is a graph stream of some underlying graph $G$ and $k$ is an integer. 
The goal is to construct a $k$-matching in $G$ (if $G$ is an unweighted graph) or a maximum weighted 
$k$-matching in $G$ (if $G$ is a weighted graph). We will consider the problem in both the insert-only 
and the dynamic streaming models. We formally define the problems under consideration: 
\begin{quote}
\umatch\\
{\bf Given:} \ a graph stream $\SSS$ for an unweighted graph $G$ and an integer $k$,\\
{\bf Goal:} \hspace*{2.5mm} construct a $k$-matching in $G$ or report that no $k$-matching exists in $G$.
  
\medskip

\wmatcheq \\
{\bf Given:} \ a graph stream $\SSS$ for a weighted graph $G$ and an integer $k$,\\
{\bf Goal:} \hspace*{2.5mm} construct a maximum weighted $k$-matching in $G$ or report that no\\
\hspace*{14.5mm} $k$-matching exists in $G$.
\end{quote} 

We will design streaming algorithms for the above problems. 
 
\subsection{Probability} 

For any probabilistic events $E_1, \ldots, E_r$, the \emph{union bound} states that 
$\Pr[\bigcup_{i=1}^r E_i] \leq \sum_{i=1}^r \Pr[E_i]$. For any random variables $X_1, \ldots, X_r$ whose 
expectations are well-defined, the {\it linearity of expectation} states that 
${\it Exp}[\sum_{i=1}^r X_i] = \sum_{i=1}^r {\it Exp}[X_i]$. A set of discrete random variables 
$\{X_1, \ldots, X_j\}$ is \emph {$\lambda$-wise independent} if for any subset $J \subseteq \{1, \ldots, j\}$ 
with $|J| \le \lambda$ and for any values $x_i$, $i \in J$, we have 
$\Pr[\bigwedge_{i\in J} X_i=x_i] = \prod_{i\in J} \Pr[X_i=x_i]$. A random variable is a \emph{$0$-$1$ 
random variable} if it only takes the values $0$ and $1$. The following theorem bounds the tail probability 
of the sum of $0$-$1$ random variables with limited independence:

\begin{proposition} [Theorem 2 in \cite{Chernoff1995}]
\label{chernoff-bound}
Given a set  $\{ X_1, \ldots, X_j \}$ of $0$-$1$ random variables, let $X=\sum_{i=1}^j X_i$ and 
$\mu = {\it Exp}[X]$. For any $\delta > 0$, if $\{ X_1, \ldots, X_j \}$ is $\lceil \mu\delta \rceil$-wise 
independent, then 
\begin{equation}
 \Pr[X \ge \mu(1+\delta)] \leq 
  \begin{cases}
    e^{-\mu \delta^2/3}       & \quad \text{if } \delta <1 \\
    e^{-\mu \delta/3}     & \quad \text{if } \delta \ge 1   \nonumber
  \end{cases}
\end{equation}
\end{proposition}

\subsection{$\ell_0$-samplers} 
\label{subsec:sampler} 

Let $0<\delta<1$ be a parameter. Let $\SSS = (i_1, \Delta_1), (i_2, \Delta_2), \ldots, (i_p, \Delta_p), \ldots$ 
be a stream of updates of an underlying vector $\textbf{x} \in \mathbb{R}^n$, where for each $j$, $i_j \in [n]$,  
$\Delta_j \in \mathbb{R}$, and the update $(i_j, \Delta_j)$ updates the $i_j$-th coordinate of $\textbf{x}$ 
by setting $\textbf{x}_{i_j} = \textbf{x}_{i_j} + \Delta_j$. An {\it $\ell_0$-sampler} for $\textbf{x} \neq 0$ either 
fails with probability at most $\delta$, or conditioned on not failing, returns a pair $(j, \textbf{x}_j)$ with 
probability $1/||\textbf{x}||_0$ for any non-zero coordinate $\textbf{x}_j$ of $\textbf{x}$, where $||\textbf{x}||_0$ 
is the $\ell_0$-norm of $\textbf{x}$, which is the number of non-zero coordinates of $\textbf{x}$. For more 
details, we refer to \cite{l0-sampling}.

Based on the results in~\cite{l0-sampling,pods11}, and as shown in~\cite{Chitnis2016}, we can develop a 
sketch-based $\ell_0$-sampler algorithm for a dynamic graph stream that samples an edge from the 
stream. More specifically, the following result was shown in~\cite{Chitnis2016}:

\begin{proposition}[Proof of Theorem~2.1 in~\cite{Chitnis2016}]
\label{lem:sampler}
Let $0 < \delta < 1$ be a parameter. There exists an $\ell_0$-sampler algorithm that, given a dynamic 
graph stream, either returns {\sc fail} with probability at most $\delta$, or returns an edge chosen 
\emph{u.a.r.}~amongst the edges of the stream that have been inserted and not deleted. This 
$\ell_0$-sampler algorithm can be implemented using $O(\log^2{n} \cdot \log(\delta^{-1}))$ bits 
of space and $O(\plog)$ update time, where $n$ is the number of vertices of the graph stream.
\end{proposition}

\subsection{Hash functions}

Let $U$ be a finite set of $n$ elements (without loss of generality, we will assume that $U = [n]^-$). A hash 
function $h$ from $U$ is \emph{perfect} w.r.t.~a subset $S$ of $U$ if it is injective on $S$, i.e., 
$h(x) \neq h(y)$ for any two distinct $x$ and $y$ in $S$. For a family $\mathcal{H}$ of hash functions, 
we write $h \xleftarrow{u.a.r.} \mathcal{H}$ to denote that the hash function $h$ is chosen 
\emph{u.a.r.}~from $\mathcal{H}$. 

Let $r$ be an integer, $0 < r \leq |U|$. A family $\mathcal{H}$ of hash functions, each mapping $U$ to 
$[r]^-$, is called \emph{universal} if for each pair of distinct elements $x, y \in U$, the number of hash 
functions $h\in \mathcal{H}$ for which $h(x)=h(y)$ is at most $|\mathcal{H}|/r$, or equivalently, for a 
hash function $h \xleftarrow{u.a.r.} \mathcal{H}$, we have $\Pr[h(x) = h(y)] \le1/r$.

The following universal family of hash functions has been well-known (see Chapter 11, \cite{Cormen}): 

\begin{proposition} 
\label{ceq1}
The collection ${\cal H}=\{h_{a,b,r} \mid 1 \le a \le p-1, 0 \le b \le p-1\}$ is a universal family of 
hash functions from $U$ to $[r]^-$, where $p \geq |U|$ is a prime number, and $h_{a,b,r}$ is defined 
as $h_{a,b,r}(x)=((ax+b) \pmod p) \pmod r$.
\end{proposition}

A perfect hash function can be retrieved from a universal family of hash functions, as given in the 
following proposition.

\begin{proposition} [Theorem 11.9 in \cite{Cormen}] 
\label{lemma-hash}
Let $\mathcal{H}$ be a universal family of hash functions, each mapping the finite set $U$ to $[r^2]^-$. 
For any set $S$ of $r$ elements in $U$, the probability that a hash function $h \xleftarrow{u.a.r.} \mathcal{H}$ 
is perfect w.r.t.~$S$ is larger than $1/2$. 
\end{proposition}

A family $\mathcal{H}$ of hash functions mapping $U$ to $[r]^-$ is {\it $\kappa$-wise independent} if for 
any $\kappa$ distinct elements $x_1, x_2, \ldots, x_{\kappa}$ in $U$, and any (not necessarily distinct) 
$a_1, a_2, \ldots, a_{\kappa}$ in $[r]^-$, we have 
\[ 
    \Pr_{h \xleftarrow{u.a.r.} \mathcal{H}} [ (h(x_1) = a_1) \wedge (h(x_2) = a_2)
      \wedge \cdots \wedge (h(x_{\kappa}) =  a_{\kappa}) ] = 1/r^{\kappa}.
\]
 
\begin{proposition}[Corollary 3.34 in \cite{Salil}]  
\label{evaluate-time}
For each integer $\kappa > 0$, there is a family of $\kappa$-wise independent functions 
$\mathcal{H}= \{ h: U \rightarrow [r]^-\}$ such that choosing a random function $h$ from $\mathcal{H}$ 
takes space $O(\kappa \log |U|)$. Moreover, evaluating a function $h$ from $\mathcal{H}$ on 
an element $x$ in $U$, i.e., computing the value $h(x)$, takes time polynomial in $\kappa$ and 
$\log |U|$.\footnote{In the original statement of this theorem in \cite{Salil}, the hash functions $h$ 
in the family $\cal H$ map binary strings of length $n$ to binary strings of length $m$ for some 
fixed integers $n$ and $m$. Here we have simplified the expressions. Thus, by $j = h(x)$, we really 
mean that $j$ is the integer whose binary representation $j_{\it bin}$ is the result of $h(x_{\it bin})$, 
where $x_{\it bin}$ is the binary representation of $x$.} 
\end{proposition}

\section{Streaming algorithms on the insert-only model}
\label{sec:insert-only}

In this section, we give a streaming algorithm for \wmatcheq{}, and hence for \umatch{} 
as a special case, in the insert-only model. We start with some notations. 

Let $G = (V, E)$ be a weighted graph with a weight function $wt: E \rightarrow \mathbb{R}_{\geq 0}$, 
where $V = [n]^-$. We define a new function $\beta: E \rightarrow \mathbb{R}_{\geq 0} \times V \times V$ 
that on an edge $e = [u, v]$ in $G$, where $u < v$, $\beta(e)=(wt(e), u, v)$. Observe that each edge in 
$G$ has a distinct $\beta$-value, and that the lexicographic order w.r.t.~$\beta$ defines a total order 
of the edges in $G$. The {\it $i$-th heaviest edge} in an edge set $E'$ is the edge that has the $i$-th 
largest $\beta$-value among all edges in $E'$. Because each edge has a distinct $\beta$-value, the 
$i$-th heaviest edge in $E'$ is uniquely defined. Note that the ``heaviness'' of edges is defined in terms 
of the edge $\beta$-values, while the ``weight'' of edges, which is used to measure the weight of matchings 
in the graph, is defined in terms of the original edge weight function $wt$ of the graph.
 
Let $f: V \rightarrow [4k^2]^-$ be a hash function. The function $f$ partitions the vertex set $V$ of $G$ 
into a collection of subsets ${\cal V}= \{V_0, V_1, \ldots, V_{4k^2-1}\}$, where for each $i \in [4k^2]^-$, 
the subset $V_i$ consists of the vertices $v$ in $V$ such that $f(v) = i$. A matching $M$ in $G$ is said 
to be \emph{nice} w.r.t.~$f$ if no two vertices of $M$ belong to the same subset $V_i$ for any $i$. If the 
hash function $f$ is clear from the context, we will simply say that $M$ is nice. 

For a subgraph $H$ of the graph $G$, we define the \emph{compact subgraph} of $H$ under $f$, 
denoted $\CC_f(H)$, as the subgraph of $H$ consisting of the edges $e$ of $H$ such that the two 
endpoints of $e$ belong to two distinct subsets $V_i$ and $V_j$ in the collection $\cal V$, and that 
$\beta(e)$ is maximum over all edges between $V_i$ and $V_j$ in $H$. 

Furthermore, we define the \emph{reduced compact subgraph} of $H$ under $f$, denoted $\RC_f(H)$, 
obtained from the compact subgraph $\CC_f(H)$ using the following procedure:  

\medskip

(1) Delete the edges $[u, v]$ in $\CC_f(H)$, where $u \in V_i$ and $v \in V_j$ for some $i \neq j$,
      if $[u, v]$ is\\
\hspace*{11mm} either not among the $2k$ heaviest edges incident to vertices in $V_i$ or not 
      among the\\
\hspace*{11mm} $2k$ heaviest edges incident to vertices in $V_j$. Let the resulting graph be 
     $\RC_f'(H)$.

\smallskip 

(2) Delete all edges $e$ in $\RC_f'(H)$ if $e$ is not among the  $4k^2$ heaviest edges in 
     $\RC_f'(H)$.

\medskip
 
The resulting graph is the {\it reduced compact subgraph $\RC_f(H)$}. 

\begin{lemma} 
\label{lemma:reduced}
The compact subgraph $\CC_f(H)$ has nice $k$-matchings if and only if the reduced compact 
subgraph $\RC_f(H)$ has nice $k$-matchings. If this is the case, then the weight of a maximum 
weighted nice $k$-matching in $\CC_f(H)$ is equal to that in $\RC_f(H)$. 

\begin{proof}
From the definition, the reduced compact subgraph $\RC_f(H)$ is a subgraph of the compact subgraph 
$\CC_f(H)$. Thus, every nice $k$-matching in $\RC_f(H)$ is also a nice $k$-matching in $\CC_f(H)$. 
In particular, the weight of a maximum weighted nice $k$-matching in $\RC_f(H)$ cannot be larger 
than that in $\CC_f(H)$. 

For the other direction, suppose that $\CC_f(H)$ has nice $k$-matchings. For convenience, we will say 
that an edge $e$ is ``incident'' to a subset $V_i$ if an endpoint of $e$ is in $V_i$, and that a $k$-matching 
$M$ ``covers'' a subset $V_i$ if $M$ has an edge incident to $V_i$.  Let $M_c$ be a maximum weighted 
nice $k$-matching in $\CC_f(H)$ that contains the largest number of edges in  $\RC_f'(H)$, which is 
the graph given in the construction of the reduced compact subgraph $\RC_f(H)$ from the compact 
subgraph $\CC_f(H)$. We prove that $M_c$ is entirely contained in the graph $\RC_f'(H)$. 

To the contrary, suppose that there is an edge $e$ in the matching $M_c$ in $\CC_f(H)$ that is not in 
$\RC_f'(H)$. Then $e$ is incident to a subset $V_{i_1}$ but is not among the $2k$ heaviest edges 
incident to $V_{i_1}$ in the graph $\CC_f(H)$. Thus, there are more than $2k$ edges incident to 
$V_{i_1}$ in the graph $\CC_f(H)$. Since the nice $k$-matching $M_c$ covers only $2k$ subsets 
in the collection $\cal V$, there must be an edge $e_1$ in $\CC_f(H)$ among the $2k$ heaviest edges 
incident to $V_{i_1}$ whose other end is incident to a subset $V_{i_2}$ not covered by $M_c$. Note 
that we have $\beta(e) < \beta(e_1)$. The edge $e_1$ cannot be among the $2k$ heaviest edges incident 
to $V_{i_2}$ --- otherwise, $e_1$ would be in $\RC_f'(H)$ and $M_c \setminus \{e\} \cup \{e_1\}$ would 
be a maximum weighted $k$-matching in $\CC_f(H)$ that contains more edges in $\RC_f'(H)$ than 
$M_c$ does, contradicting the assumption of $M_c$. Using the same argument on the subset $V_{i_2}$, 
we can find an edge $e_2$ among the $2k$ heaviest edges incident to $V_{i_2}$ whose other endpoint 
is in a subset $V_{i_3}$ not covered by $M_c$ such that $e_2$ is not among the $2k$ heaviest edges 
incident to $V_{i_3}$ and that $\beta(e_1) < \beta(e_2)$. Continuing this process will produce a sequence 
of edges $e_1$, $e_2$, $\cdots$, with $\beta(e_h) < \beta(e_{h+1})$ for all $h$. Since the graph $\CC_f(H)$ 
is finite, we must have some edges repeating in this sequence, i.e., we will find $e_s$ and $e_t$ in the 
sequence with $s < t$ such that $e_s = e_t$. However, this would imply that  
\[ \beta(e_s) < \beta(e_{s+1}) < \cdots < \beta(e_{t-1}) < \beta(e_t) = \beta(e_s), \]
which is impossible. This contradiction shows that the maximum weighted nice $k$-matching $M_c$ 
is entirely contained in the graph $\RC_f'(H)$. As a consequence, $M_c$ is a maximum weighted 
nice $k$-matching in $\RC_f'(H)$.

If $\RC_f'(H)$ contains no more than $4k^2$ edges, then $\RC_f'(H) = \RC_f(H)$ and $M_c$ 
is also a maximum weighted nice $k$-matching in $\RC_f(H)$. 

If $\RC_f'(H)$ contains more than $4k^2$ edges, then the reduced compact subgraph $\RC_f(H)$ 
contains exactly the $4k^2$ heaviest edges in $\RC_f'(H)$. Now consider a maximum weighted 
$k$-matching $M_c'$ in $\RC_f'(H)$ that contains the maximum number of edges in the reduced 
compact subgraph $\RC_f(H)$. If there is an edge $e'$ in $M_c'$ that is not in $\RC_f(H)$, then delete all 
vertices in $\RC_f(H)$ that are in the subsets of $\cal V$ covered by $M_c' \setminus \{e'\}$. This will 
delete no more than $2(k-1) \cdot 2k < 4k^2$ edges in $\RC_f(H)$, because each subset $V_i$ is 
incident to at most $2k$ edges in $\RC_f(H)$. Therefore, there is at least one edge $e''$ left in $\RC_f(H)$, 
and by definition, $\beta(e'') > \beta(e')$. Thus, the matching $M_c' \setminus \{e'\} \cup \{e''\}$ would be 
a maximum weighted nice $k$-matching in $\RC_f'(H)$ that contains more edges in $\RC_f(H)$ than 
$M_c'$ does, but this contradicts the assumption of the matching $M_c'$. This contradiction shows that 
the maximum weighted nice $k$-matching $M_c'$ in $\RC_f'(H)$ is also a maximum weighted nice 
$k$-matching in $\RC_f(H)$. This completes the proof. 
\end{proof}
\end{lemma}

The following lemma shows how we construct the reduced compact subgraph $\RC_f(H)$ when 
the subgraph $H$ has been stored in memory.

\begin{lemma}
\label{lemma:rc}
Let $f$ be a hash function mapping the vertex set of a graph $G$ to $[4k^2]^-$. There is an algorithm 
that, for any subgraph $H$ of $G$, constructs the subgraph $\RC_f(H)$ in time and space both bounded 
by $O(|H|+k^2)$. 

\begin{proof}
For each $i \in [4k^2]^-$, remember that $V_i$ is the subset of vertices in $G$ whose image under $f$ is $i$.  
The algorithm on a subgraph $H$ of $G$ works as follows. By going through the edges of the  
graph $H$ and using the hash function $f$, the algorithm deletes, in time $O(|H|+k^2)$, all edges 
whose two endpoints have the same image under $f$, and puts each $[u, v]$ of the remaining edges in 
the sets $E_i$ and $E_j$, where $i = f(u)$ and $j = f(v)$. Now for each set $E_i$, the algorithm sorts,  
using Radix-Sort in linear time, the edges $[u, v]$ in $E_i$ in terms of the pairs $(f(u), f(v))$, identifies  
the heaviest edge (i.e., the edge with the maximum $\beta$-value) between the sets $V_i$ and $V_j$, 
for each $j$, and deletes all other edges between $V_i$ and $V_j$. The result of this process is the 
compact subgraph $\CC_f(H)$. Now using the linear-time selection algorithm \cite{Cormen} on each set 
$E_i$, the algorithm can identify the $(2k)$-th heaviest edge in the set $E_i$, and, by going through all 
edges in $\CC_f(H)$, delete the edges $[u, v]$ between $E_i$ and $E_j$, for all $i, j \in [4k^2]^-$, if $[u, v]$ 
is either not among the $2k$ heaviest edges in $E_i$ or not among the $2k$ heaviest edges in $E_j$. 
This gives the subgraph $\RC_f'(H)$. Finally, using the linear-time selection algorithm one more time, the 
algorithm can delete the edges in $\RC_f'(H)$ that are not among the $4k^2$ heaviest, and obtain the 
reduced compact subgraph $\RC_f(H)$. This shows that the running time of the algorithm is bounded by 
$O(|H|+k^2)$, which also bounds the space complexity of the algorithm. 
\end{proof}
\end{lemma}

We now describe our streaming algorithm in the insert-only model for the \wmatcheq{} problem. Let 
$(\SSS, k)$ be an instance of the problem, where $\SSS = \{ (e_1, wt(e_1)), \ldots, (e_s, wt(e_s)), \ldots \}$ 
is the stream of inserting edges of a graph $G$. For each $s$, let $G_s$ be the subgraph of $G$ consisting 
of the first $s$ edges $e_1, \ldots, e_s$ of $\SSS$, and for $r \le s$, let $G_{r, s}$ be the subgraph of 
$G$ consisting of the edges $e_r, e_{r+1}, \ldots, e_s$. Let $f$ be a hash function mapping the vertex set 
of the graph $G$ to $[4k^2]^-$. For each integer $s$, we denote by $\hat{s}$ the largest multiple of $4k^2$ 
that is strictly smaller than $s$, i.e., $\hat{s} = 4k^2 \cdot i$ for an integer $i$ and $s = \hat{s} + q$ with 
$1 \leq q \leq 4k^2$. Note that even when $s$ is a multiple of $4k^2$, we still have $\hat{s} < s$.

Let $G_0^f = \emptyset$. For each $s > 0$, we define, recursively, a subgraph $G_s^f$ of the graph 
$G_s$ as 
\begin{equation}
 G_s^f = \RC_f(G_{\hat{s}}^f \cup G_{\hat{s}+1,s}).     \label{chen-eq1}
\end{equation}

\begin{lemma}
\label{chenlem1}
For each $s \geq 0$, the graph $G_s^f$ has at most $4k^2$ edges. Moreover, $G_s^f = \RC_f(G_s)$. 

\begin{proof}
The bound on the size of the graph $G_s^f$ comes directly from the definition of the reduced compact 
subgraphs under the hash function $f$.

To prove the equality $G_s^f = \RC_f(G_s)$, we first prove the following claim:
\begin{quote}
{\bf Claim.} Let $H_1$ be a subgraph of $H_2$ and let $e$ be an edge in $H_1$. Then $e \in \RC_f(H_2)$ 
implies $e \in \RC_f(H_1)$.
\end{quote}

Consider the edge $e$ in $H_1$ that is in $\RC_f(H_2)$. Then (1) $e$ is the heaviest edge in $H_2$ 
between two vertex sets $V_i$ and $V_j$ in the partition by the hash function $f$ (i.e., $e$ is in the 
compact subgraph $\CC_f(H_2)$); and (2) $e$ is among the $2k$ heaviest edges incident to $V_i$ and 
among the $2k$ heaviest edges incident to $V_j$ in the graph $\CC_f(H_2)$ (i.e., $e$ is in the graph 
$\RC_f'(H_2)$); and (3) $e$ is among the $4k^2$ heaviest edges in $\RC_f'(H_2)$. Since $H_1$ 
is a subgraph of $H_2$ and $e \in H_1$, by the conditions (1)-(3) above, it is easy to verify that (1') $e$ 
is in the compact subgraph $\CC_f(H_1)$; (2') $e$ is in the graph $\RC_f'(H_1)$; and (3') $e$ is 
among the $4k^2$ heaviest edges in $\RC_f'(H_1)$, i.e., the edge $e$ must be in the reduced compact 
subgraph $\RC_f(H_1)$. This proves the claim. 

\medskip

Now we get back to the proof of $G_s^f = \RC_f(G_s)$. Our proof goes by induction on $s \geq 0$. The 
equality obviously holds true for $s \leq 4k^2$, since in this case $\hat{s} = 0$ so $G_{\hat{s}} = \emptyset$. 
Thus, we will assume $s > 4k^2$. Note that $G_s = G_{\hat{s}} \cup G_{\hat{s}+1,s}$.

Let $e$ be an edge in $\RC_f(G_s)$. If $e$ is in $G_{\hat{s}} \setminus G_{\hat{s}}^f$. Then since 
$e \in \RC_f(G_s)$ and  $G_{\hat{s}}$ is a subgraph $G_s$, by the Claim we proved above, we would have 
$e \in \RC_f(G_{\hat{s}}) = G_{\hat{s}}^f$, contradicting the assumed condition. Thus, this case is not 
possible, and we must have $e \in G_{\hat{s}}^f \cup G_{\hat{s}+1,s}$. Since 
$G_{\hat{s}}^f \cup G_{\hat{s}+1,s}$ is a subgraph of $G_s$ and $e \in \RC_f(G_s)$, by the Claim above, 
$e \in \RC_f(G_{\hat{s}}^f \cup G_{\hat{s}+1,s}) = G_s^f$. Since $e$ is an arbitrary edge in $\RC_f(G_s)$,  
this proves that $\RC_f(G_s)$ is a subgraph of $G_s^f$.

For the other direction, let $e$ be an edge in 
$G_s^f = \RC_f(G_{\hat{s}}^f \cup G_{\hat{s}+1,s}) = \RC_f(\RC_f(G_{\hat{s}}) \cup G_{\hat{s}+1,s})$, 
where the second equality is by the inductive hypothesis, and assume that $e$ is an edge between two 
vertex sets $V_i$ and $V_j$ in the vertex partition by the hash function $f$. Then, (1) $e$ is the heaviest 
edge between $V_i$ and $V_j$ in the graph $\RC_f(G_{\hat{s}}) \cup G_{\hat{s}+1,s}$, thus, is also such 
an edge in the graph $G_{\hat{s}} \cup G_{\hat{s}+1,s} = G_s$, i.e., $e$ is an edge in $\CC_f(G_s)$; (2) 
$e$ is among the $2k$ heaviest edges incident to $V_i$ in the graph $\CC_f(G_s)$: if not, then there 
must be an edge $e'$ incident to $V_i$ in $\CC_f(G_s)$ with $\beta(e') > \beta(e)$ and 
$e' \not\in \RC_f'(\RC_f(G_{\hat{s}}) \cup G_{\hat{s}+1,s})$. However, this is impossible because $e'$ 
must be in $\CC_f(\RC_f(G_{\hat{s}}) \cup G_{\hat{s}+1,s})$ and the edge $e$ is among the $2k$ 
heaviest edges incident to $V_i$ in $\CC_f(\RC_f(G_{\hat{s}}) \cup G_{\hat{s}+1,s})$. Similarly, $e$ is 
among the $2k$ heaviest edges incident to $V_j$ in $\CC_f(G_s)$. Thus, the edge $e$ is in $\RC_f'(G_s)$; 
and (3) $e$ must be among the $4k^2$ heaviest edges in $\RC_f'(G_s)$: if not, there must be an edge $e''$ 
in $\RC_f(G_s)$ with $\beta(e'') > \beta(e)$ and $e'' \not\in \RC_f(G_{\hat{s}}^f \cup G_{\hat{s}+1,s}))$. 
By the Claim we proved above, $e''$ cannot be in $G_{\hat{s}}^f \cup G_{\hat{s}+1,s}$. The remaining 
possibility is $e'' \in G_{\hat{s}} \setminus G_{\hat{s}}^f$, but this, plus $e'' \in \RC_f(G_s)$ and the 
Claim above would give the contradiction $e'' \in \RC_f(G_{\hat{s}}) = G_{\hat{s}}^f$. Therefore, the edge 
$e$ must be in the graph $\RC_f(G_s)$, proving that $G_s^f$ is a subgraph of $\RC_f(G_s)$. 

Combining the above two cases proves that $G_s^f = \RC_f(G_s)$.
\end{proof}
\end{lemma}

Now we are ready to present the streaming algorithm  {\bf w-Match}$_{\it ins}$ in the insert-only 
model for the \wmatcheq{} problem, as given in Figure~\ref{chenfig1}, where 
$\SSS = \{ e_1, e_2 \ldots, e_s, \ldots \}$ is an edge stream of a graph $G$ (here we assume that 
the edge weight has been included in each edge $e_s$ in the stream), $k$ is the parameter, and 
$\epsilon > 0$ is a fixed constant that bounds the error probability. Note that the algorithm has used 
some notations that are used in Lemma~\ref{chenlem1} and its proof. 

\begin{figure}[htbp]
\setbox4=\vbox{\hsize31pc
\medskip
 \noindent\strut  
\hspace*{3mm}\footnotesize {\bf Algorithm {\bf w-Match}$_{\it ins}(\SSS, k)$}\\
\hspace*{3mm}{\sc input}:  an edge stream $\SSS = \{ e_1, e_2, \ldots, e_s, \ldots \}$ 
    of a weighted graph $G$ and parameter $k$\\
\hspace*{3mm}{\sc output}: a maximum weighted $k$-matching in $G$, or report that no 
   $k$-matching is in $G$. 

\smallskip

\hspace*{3mm}$\backslash\!\backslash$ Preprocessing\\
\hspace*{3mm}1. \ let ${\cal H}_{\epsilon}$ be a set of $\lceil \log (1/\epsilon) \rceil$ hash functions 
    picked {\it u.a.r.}~from a universal family of\\
\hspace*{7mm} hashing functions mapping $V(G)$ to $[4k^2]^-$;\\
\hspace*{3mm}2. \ {\bf for} (each hash function $f$ in ${\cal H}_{\epsilon}$) \ $G_0^f = \emptyset$;\\
\hspace*{3mm}3. \  $G_{1,s} = \{e_1, e_2, \ldots, e_s\}$, where either $s = 4k^2$, or $s < 4k^2$ and 
   $e_s$ is the last edge in $\SSS$;

\smallskip

\hspace*{3mm}$\backslash\!\backslash$ Updating\\
\hspace*{3mm}4. \ {\bf while} (the edge stream $\SSS$ is not ended) \\
\hspace*{13mm} interleave the executions of steps 4.1 and 4.2 so that the time between reading\\ 
\hspace*{13mm} two consecutive edges in $\SSS$ into $G_{s+1, s'}$ is bounded by $O(1)$: \\
\hspace*{3mm}4.1 \hspace*{5mm} {\bf for} (each hash function $f$ in ${\cal H}_{\epsilon}$) 
     \  $G_s^f = \RC_f(G_{\hat{s}}^f \cup G_{\hat{s}+1,s})$; \\
\hspace*{3mm}4.2 \hspace*{5mm} 
  $G_{s+1, s'} = \{ e_{s+1}, e_{s+2}, \ldots, e_{s'} \}$,\\ 
\hspace*{13mm} where either $s' = s+4k^2$, or $s' < s+4k^2$ 
    and $e_{s'}$ is the last edge in $\SSS$;\\
\hspace*{3mm}4.3 \hspace*{5mm}  $s = s'$;

\smallskip

\hspace*{3mm}$\backslash\!\backslash$ Query\\
\hspace*{3mm}5. \ \hspace*{1mm} {\bf for} (each hash function $f$ in ${\cal H}_{\epsilon}$)\\
\hspace*{3mm}5.1 \hspace*{5mm} $G_f = \RC_f(G_{\hat{s}}^f \cup G_{\hat{s}+1,s})$;\\
\hspace*{3mm}5.2 \hspace*{5mm} construct the maximum weighted $k$-matching in $G_f$;\\
\hspace*{3mm}6. \ \hspace*{1mm} {\bf if} (a $k$-matching is constructed in step 5)\\
\hspace*{9mm} {\bf then} return the $k$-matching with the largest weight constructed in step 5;\\
\hspace*{9mm} {\bf else} return(''no $k$-matching exists in $G$").
\medskip
\strut} $$\boxit{\box4}$$
 \vspace{-8mm}
\caption{A streaming algorithm for \wmatcheq{} in the insert-only model} 
\label{chenfig1}
\end{figure}

\begin{lemma} 
\label{lem16}
The algorithm {\bf w-Match}$_{\it ins}(\SSS, k)$ runs in space $O(k^2)$ and has update time 
$O(1)$. 

\begin{proof}
Since $\epsilon > 0$ is a fixed constant, the number of hash functions in the set ${\cal H}_{\epsilon}$ 
is a constant. Thus, steps 1-2 of the algorithm {\bf w-Match}$_{\it ins}(\SSS, k)$ take constant 
time and need constant space to store the hash functions in ${\cal H}_{\epsilon}$.

Step 3 of the algorithm uses $O(k^2)$ space to store the $s \leq 4k^2$ edges in the subgraph 
$G_{1,s}$.  

Step 4 of the algorithm works with four subgraphs: $G_{\hat{s}}^f$, $G_{\hat{s}+1,s}$, $G_s^f$, 
and $G_{s+1,s'}$, where $s' \leq s + 4k^2$. By definition, each of $G_{\hat{s}+1, s}$ and $G_{s+1,s'}$ 
contains at most $4k^2$ edges, and, by Lemma~\ref{chenlem1}, each of the subgraphs 
$G_{\hat{s}}^f$ and $G_s^f$ contains at most $4k^2$ edges. Moreover, since the size of the graph 
$G_{\hat{s}}^f \cup G_{\hat{s}+1,s}$ is bounded by $O(k^2)$, by Lemma~\ref{lemma:rc}, the 
computation of step 4.1 takes space $O(k^2)$. In summary, step 4 of the algorithm runs in 
space $O(k^2)$.

Similarly, for each hash function $f$ in ${\cal H}_{\epsilon}$, step 5.1 takes space $O(k^2)$. 
Since the size of the graph $G_f$ is bounded by $O(k^2)$, step 5.2 also takes space $O(k^2)$. 
Since the number of hash functions in ${\cal H}_{\epsilon}$ is a constant, we conclude that step 5 of 
the algorithm runs in space $O(k^2)$.

This shows that the algorithm {\bf w-Match}$_{\it ins}(\SSS, k)$ runs in space $O(k^2)$. Now we 
consider the update time of the algorithm. The update time for reading each of the first $s \leq 4k^2$  
edges in the stream $\SSS$ in step 3 is obviously $O(1)$. The rest of the edges in the stream $\SSS$ 
are read in step 4. Since the size of the graph $G_{\hat{s}}^f \cup G_{\hat{s}+1,s}$ is bounded by 
$O(k^2)$, by Lemma~\ref{lemma:rc}, step 4.1 of the algorithm runs in time $O(k^2)$. Therefore, the 
execution of step 4.1 can be divided into $4k^2$ segments such that each segment takes time $O(1)$. 
Now the interleaved execution of steps 4.1 and 4.2 can read an edge in the stream $\SSS$ in step 4.2 
with the execution of a segment of step 4.1, until either the set $G_{s+1,s'}$ has $4k^2$ edges or the 
stream end is encountered. This guarantees an update time of $O(1)$  for reading each of the edges in  
the input stream $\SSS$.   
\end{proof}
\end{lemma}

We conclude this section with the following theorem. 

\begin{theorem}
\label{thm:insert}
For any fixed $\epsilon > 0$, the streaming algorithm {\bf w-Match}$_{\it ins}(\SSS, k)$, where $\SSS$ 
is a stream for a weighted graph $G$ in the insert-only model, runs in space $O(k^2)$ and update time 
$O(1)$, and (1) if $G$ has $k$-matchings then the algorithm returns a maximum weighted $k$-matching 
in $G$ with probability $\geq 1 - \epsilon$; and (2) if $G$ has no $k$-matchings then the algorithm 
reports so. 

\begin{proof}
First note that for each hash function $f$ in ${\cal H}_{\epsilon}$, the graph $G_f$ constructed in 
step 5.1 is a subgraph of the graph $G$. Therefore, if the graph $G$ has no $k$-matchings, then 
the graph $G_f$ cannot have $k$-matchings. Thus, in this case, the algorithm reports correctly. 

Now assume that the graph $G$ has $k$-matchings. Let $M$ be a maximum weighted $k$-matching 
in $G$. The vertex set $V(M)$ of the matching $M$ has $2k$ vertices. By Proposition~\ref{lemma-hash}, 
each hash function $f$ in ${\cal H}_{\epsilon}$, which is {\it u.a.r.}~picked from a universal family of 
hash functions mapping $V(G)$ to $[(2k)^2]^- = [4k^2]^-$, has probability at least $1/2$ to be perfect 
w.r.t.~$V(M)$. Since there are $\lceil \log(1/\epsilon) \rceil$ hash functions in ${\cal H}_{\epsilon}$, with 
probability at least $1 - 1/2^{\lceil \log(1/\epsilon) \rceil} \geq 1 - \epsilon$, there is a hash function $f_0$ 
in ${\cal H}_{\epsilon}$ that is perfect w.r.t.~$V(M)$. As a result, the maximum weighted $k$-matching 
$M$ in $G$ is a maximum weighted nice $k$-matching in the compact subgraph $\CC_{f_0}(G)$. 
By Lemma~\ref{lemma:reduced}, the reduced compact subgraph $\RC_{f_0}(G)$ also has a 
$k$-matching $M_0$ whose weight is equal to that of $M$. Since $\RC_{f_0}(G)$ is a subgraph of 
$G$, $M_0$ is also a maximum weighted $k$-matching in $G$. 

By Lemma~\ref{chenlem1} and step 5.1 of the algorithm, we have $G_{f_0} = \RC_{f_0}(G)$. Therefore, 
the matching $M_0$ is a (maximum weighted) $k$-matching in the graph $G_{f_0}$, and the maximum 
weighted $k$-matching constructed in step 5.2 for the graph $G_{f_0}$, which could be different from 
$M_0$ but must have the same weight as $M_0$, is a maximum weighted $k$-matching in the graph 
$G$, which will be returned in step 6 of the algorithm. This completes the proof that if the graph $G$ 
has $k$-matchings, then with probability at least $1 - \epsilon$, the algorithm 
{\bf w-Match}$_{\it ins}(\SSS, k)$ returns a maximum weighted $k$-matching in the graph $G$.
\end{proof}
\end{theorem}

Note that the algorithm {\bf w-Match}$_{\it ins}(\SSS, k)$ given in Figure~\ref{chenfig1} queries and 
computes a maximum weighted $k$-matching at the end of the stream. It is easy to see that the 
algorithm can be trivially modified so that it can query and compute a maximum weighted $k$-matching 
in the graph $G_s$ consisting of the first $s$ edges in the stream for any $s \geq 0$ after seeing the 
edge $e_s$, keeping the same space and update time complexities.

\section{Streaming algorithms on the dynamic model}
\label{sec:dynamic}

In this section, we present a streaming algorithm for \wmatcheq{}, thus also for \umatch{}, on the dynamic 
model.  For this, we first develop a hashing scheme that uses $O(k \cdot \plog)$ space and has a high 
success probability. The streaming algorithm on the dynamic model will use the hashing scheme and the 
$\ell_0$-sampling technique discussed in Section~\ref{sec:prelim}. 

\subsection{Perfect hashing in $O(k \cdot \plog)$ space with high probability}
\label{sec:structural}

The hashing scheme developed in this subsection will be used in our streaming algorithm on the dynamic model. 
We believe that the result should also be useful in other applications. Let $S$ be an (unknown) $k$-subset 
of a universal set $U$ and suppose that we want to {\it distinguish} the $k$ elements of $S$ by constructing 
a collection of subsets of $U$ that contains $k$ pairwise disjoint subsets, each containing exactly one element 
in $S$. For example, by Proposition~\ref{lemma-hash}, a hash function $h$ picked {\it u.a.r.}~from a universal 
family of hash functions from $U$ to $[k^2]^-$ has a probability $\geq 1/2$ to be perfect w.r.t.~$S$, and thus 
distinguishes $S$. The hash function $h$ uses $O(k^2)$ space, which is large and would directly impact the 
space complexity of our streaming algorithms. Moreover, the success probability $1/2$ of $h$ is not sufficiently 
large for our purposes. An $O(k)$-space hash function perfect w.r.t.~$S$ can be constructed using a 2-level 
hashing scheme (see \cite{Cormen}, Section 11.5), in which, however, the construction of the hash functions 
in level 2 must know the set $S$ and the hashing result on $S$ in level 1. Moreover, the scheme uses multiple 
hash functions from universal hash families, which would significantly decrease the success probability. 
 
\begin{figure}[htbp]
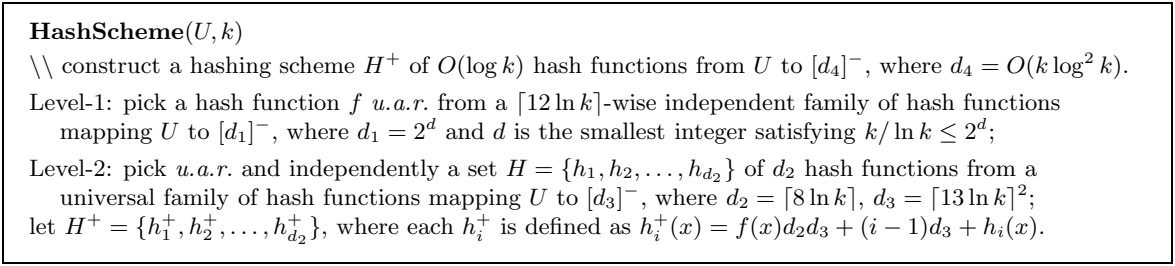


\vspace*{-2mm}

\setbox4=\vbox{\hsize36pc  \smallskip
 \noindent \strut 
\footnotesize 
\hspace*{2mm}{\bf HashScheme$(U, k)$} \\
\hspace*{2mm}$\backslash\backslash$ construct a hashing scheme $H^+$ of $O(\log k)$ hash functions from 
    $U$ to $[d_4]^-$, where $d_4 = O(k \log^2 k)$.

\smallskip

\hspace*{2mm}Level-1: pick a hash function $f$ {\it u.a.r.}~from a $\lceil 12 \ln k \rceil$-wise independent family 
    of hash functions\\
\hspace*{5mm} mapping $U$ to $[d_1]^-$, where $d_1 = 2^d$ and $d$ is the smallest integer satisfying  
    $k/\ln k \leq 2^d$;

\smallskip

\hspace*{2mm}Level-2: pick {\it u.a.r.}~and independently a set $H = \{h_1, h_2, \ldots, h_{d_2}\}$ of $d_2$ 
    hash functions from a\\
\hspace*{5mm} universal family of hash functions mapping $U$ to $[d_3]^-$, where $d_2 = \lceil 8 \ln k \rceil$, 
     $d_3 = \lceil 13 \ln k \rceil^2$;
  
\hspace*{2mm}let $H^+ = \{ h_1^+, h_2^+, \ldots, h_{d_2}^+ \}$, where each $h_i^+$ is defined as  
   $h_i^+(x) = f(x) d_2 d_3 + (i-1)d_3 + h_i(x)$.
\medskip
\strut} $$\boxit{\box4}$$
 \vspace{-10mm}
\caption{A hashing scheme that uses smaller space with higher success probability} 
\label{chenfig14}
\end{figure}

We propose a hashing scheme that follows the ideas of 2-level hashing, but with a more careful selection on 
the hashing methods and on the hashing parameters. Our scheme uses $O(k \cdot \plog)$ space but has a 
much higher success probability, and its construction in level 2 needs to know neither the set $S$ nor the 
hashing results on $S$ in level 1. The hashing scheme is given in Figure~\ref{chenfig14}. 

We give some remarks on the hashing scheme in Figure~\ref{chenfig14}. Suppose that the hash function 
$f$ in Level-1 partitions the universal set $U$ into $U_0$, $U_1$, $\ldots$, $U_{d_1-1}$. To ensure the 
pairwise disjointness of the $d_1 d_2$ sets $h_i^+(U_j)$, $1 \leq i \leq d_2$, $0 \leq j \leq d_1-1$, we define 
$h_i^+(U_j)$ to be the set $h_i(U_j)$ plus an offset $j d_2 d_3 + (i-1)d_3$. Therefore, for each $x \in U_j$, 
the function $h_i^+$ has the value 
     \[ h_i^+(x) = j d_2 d_3 + (i-1)d_3 + h_i(x) = f(x) d_2 d_3 + (i-1)d_3 + h_i(x). \]
Each $h_i^+$ is a function mapping $U$ to $[d_4]^-$, where $d_4 = d_1 d_2 d_3 = O(k \log^2 k)$. 

\begin{theorem}
\label{alg-hash-space}
The hash function set $H^+$ can be constructed in space $O(\log k \log |U|)$. For each $x \in U$ and 
each $h_i^+ \in H^+$, the value of $h_i^+(x)$ can be computed in time polynomial in $\log |U|$.

\begin{proof}
By Proposition~\ref{evaluate-time}, constructing and storing the function $f$ in Level-1, which  is picked 
{\it u.a.r.}~from a $\lceil 12\ln k \rceil$-wise independent family of hash functions from $U$ to $[d_1]^-$, uses 
space $O(\log k \log |U|)$. Each hash function $h_i$ in Level-2 picked from the universal family of hash 
functions from $U$ to $[d_3]^-$, which is of the form given in Proposition~\ref{ceq1}, takes $O(1)$ space. 
Since $|H^+| = d_2 = O(\log k)$, we conclude that the set $H^+ = \{ h_1^+, h_2^+, \ldots, h_{d_2}^+ \}$ 
of hash functions can be constructed and stored in space $O(\log k \log |U|)$.

By Proposition~\ref{evaluate-time}, computing the value $j = f(x)$ takes time polynomial in $\log |U|$ and 
$\log k$. This, plus the $O(1)$ time for computing the other parts of the function $h_i^+$, shows that the 
value of $h_i^+(x)$ can be computed in time polynomial in $\log |U|$, after noting that $k \leq |U|$. 
\end{proof}
\end{theorem}

For each value $q \in [d_4]^-$, let $H_{\it inv}^+(q)$, i.e., the ``inverse'' of   $H^+$ on $q$, be the set of such 
an element $x$ in $U$ such that $h_i^+(x) = q$ for some hash function $h_i^+$ in $H^+$.   

\begin{theorem}  
\label{alg-hash-THM4}
For any subset $S$ of $U$ with $|S| = k \geq 2$, with probability at least $1 - 4/(k^3\ln k)$, there are $k$ 
disjoint subsets $H_{\it inv}^+(q_1)$, $\ldots$, $H_{\it inv}^+(q_k)$ of $U$ such that 
$|H_{\it inv}^+(q_i) \cap S| = 1$ for all $i \in [k]$. 

\begin{proof}
Recall that the function $f$ partitions the set $U$ into $d_1$ disjoint subsets $U_0$, $U_1$, $\ldots$, 
$U_{d_1-1}$. We first show that, for each $U_j$, with a high probability, there is a hash function in the set 
$H = \{ h_1, h_2, \ldots, h_{d_2} \}$ that is perfect w.r.t.~$S \cap U_j$. 

For an element $x \in S$, and for each $j \in [d_1]^-$, let $X_{x,j}$ be the $0$-$1$ random variable such that 
$X_{x,j} = 1$ if and only if $f(x) = j$. Let $X_j = \sum_{x \in S} X_{x, j}$, which is the number of elements in 
$S$ that are hashed to $j$ by the hash function $f$. Thus, $X_j = |S \cap U_j|$. 

Since $f$ is picked {\it u.a.r.}~from a $\lceil 12 \ln k \rceil$-wise independent family of hash functions, 
the random variables $X_{x, j}$, for $x \in S$, are $\lceil 12 \ln k \rceil$-wise independent and 
$\Pr[X_{x, j}=1] = 1/d_1$. Thus, ${\it Exp}[X_j] = |S|/d_1$. Since $k/\ln k \leq d_1 < 2k/\ln k$, so 
$\ln k/2 < {\it Exp}[X_j] \leq \ln k$. Applying Proposition~\ref{chernoff-bound} with $\mu = {\it Exp}[X_j]$ 
and $\delta = 12\ln k/{\it Exp}[X_j] > 1$, we get 
\[ \Pr[X_j \ge (1+\delta) {\it Exp}[X_j] ]  \leq e^{- \delta \cdot {\it Exp}[X_j] / 3} = 1/k^4.  \]
Since ${\it Exp}[X_j] \leq \ln k$ and $\delta = 12\ln k/{\it Exp}[X_j]$, we have 
$(1 + \delta){\it Exp}[X_j] \leq 13 \ln k$. Hence, 
\[  \Pr[X_j \ge 13\ln k] \leq \Pr[X_j \ge (1+\delta)E[X_j]] \leq 1/k^4.   \] 
Let $\mathcal{E}'$ denote the event that for all $j \in [d_1]^-$, $X_j < 13 \ln k$. By the union bound, 
we have 
\[  \Pr[\mathcal{E}'] \geq 1 - d_1/k^4 \geq 1 - 2/(k^3\ln k),  \]  
where the last inequality holds since $d_1 < 2k/\ln{k}$. 

Assume the event $\mathcal{E}'$ that for all $j \in [d_1]^-$, $|S \cap U_j| < 13 \ln k$ holds for all 
$j \in [d_1]^-$. For each $j \in [d_1]^-$, let ${\cal E}_j$ be the event that the set $H$ does not contain 
a hash function perfect w.r.t.~$S \cap U_j$. Since $|S \cap U_j| < 13\ln k$, and each hash function in $H$ 
is picked independently and {\it u.a.r.}~from a universal family of hash functions from $U$ to $[d_3]^-$ with 
$d_3 = \lceil 13\ln k \rceil^2$, which, by Proposition~\ref{lemma-hash}, is perfect w.r.t.~$S \cap U_j$ with 
probability $\geq 1/2$. Since $H$ consists of $d_2 = \lceil 8\ln k \rceil$ such hash functions, we derive 
$\Pr[{\cal E}_j | \mathcal{E}'] \leq 1/2^{d_2} < 1/k^4$. Applying the union bound on all $j \in [d_1]^-$, we 
conclude that, under the event $\mathcal{E}'$, the probability that there is a $j \in [d_1]^-$ such that the set 
$H$ contains no hash function perfect w.r.t.~$S \cap U_j$ is bounded by $d_1/k^4 < 2/(k^3 \ln k)$. Now let 
$\mathcal{E}''$ be the event that, under the event $\mathcal{E}'$, for every $j \in [d_1]^-$, the set $H$ contains a 
hash function $h_{i_j}$ perfect w.r.t.~$S \cap U_j$. Then $\Pr[ \mathcal{E}'' | \mathcal{E}' ] \geq 1 - 2/(k^3 \ln k)$, 
which gives directly 
\[ \Pr[\mathcal{E}' \cap \mathcal{E}''] = 
   \Pr[\mathcal{E}'' | \mathcal{E}'] \cdot \Pr[\mathcal{E}'] 
   \geq (1 - 2/(k^3\ln k))^2 \geq 1 - 4/(k^3 \ln k). \]
Since $\mathcal{E}' \cap \mathcal{E}''$ is the event in which for every $j \in [d_1]^-$, there is a hash function 
$h_{i_j}$ in the set $H$ that is perfect w.r.t.~$S \cap U_j$, which implies immediately that the hash function 
$h_{i_j}^+$ in the set $H^+$ is perfect w.r.t.~$S \cap U_j$. Thus, under the event $\mathcal{E}' \cap \mathcal{E}''$, 
the union $Q_k$ of the pairwise disjoint sets $h_{i_0}^+(S \cap U_0)$, $\ldots$, $h_{i_{d_1-1}}^+(S \cap U_{d_1-1})$ 
contains exactly $k$ values $q_1$, $\ldots$, $q_k$ in $[d_4]^-$, such that each subset $H_{\it inv}^+(q_i)$ of 
$U$ contains an element in $S$. To see the pairwise disjointness of the subsets $H_{\it inv}^+(q_1)$, $\ldots$, 
$H_{\it inv}^+(q_k)$, observe that the $d_1 d_2$ subsets $h_i^+(U_j)$, for $i \in [d_2]$ and $j \in [d_1]^-$, are 
pairwise disjoint, so each value in $Q_k$ is in a unique subset $h_i^+(U_j)$. Thus,

(1) if $q_s$, $q_t$ in $Q_k$, $q_s \neq q_t$, are in the same $h_i^+(U_j)$, then, since no element in $U_j$ can 
have two different images under the same function $h_i^+$, the subsets $H_{\it inv}^+(q_s)$ and 
$H_{\it inv}^+(q_t)$ are disjoint;  

(2) if $q_s$ and $q_t$ in $Q_k$ are in two different $h_i^+(U_j)$ and $h_{i'}^+(U_{j'})$, respectively, then we 
must have $j \neq j'$ because by the definition of $Q_k$, if $j = j'$ then we must have $i = i' = i_j$. But this 
implies that $H_{\it inv}^+(q_s) \subseteq U_j$ and $H_{\it inv}^+(q_t) \subseteq U_{j'}$ so 
$H_{\it inv}^+(q_s)$ and $H_{\it inv}^+(q_t)$ must be disjoint.
\end{proof}
\end{theorem}

\subsection{The streaming algorithm for \wmatcheq{}}

Now we are ready for our streaming algorithm for the \wmatcheq{} problem on the dynamic model. We 
first give a high-level description of the algorithm. Let $\SSS$ be a dynamic stream of a weighted graph 
$G = (V, E)$ with a weight function $wt$, and let $k$ be the parameter. Let $M_{\max}$ be any fixed 
maximum weighted $k$-matching in $G$. We first use a hashing scheme $H^+$, as given in 
Subsection~\ref{sec:structural}, to hash the vertices of $G$ into a range $[r]^-$, where $r = O(k \log^2k)$. 
By Theorem~\ref{alg-hash-THM4}, {\it w.h.p.}~there is a set $B$ of $2k$ values in $[r]^-$ such that the 
collection $H_{\it inv}^+(B) = \{ H_{\it inv}^+(i) \mid i \in B\}$ consists of $2k$ pairwise disjoint subsets of 
$V(G)$ in which each subset contains exactly one vertex in the matching $M_{\max}$. As a result, every 
edge in $M_{\max}$ appears between two different subsets in $H_{\it inv}^+(B)$, so we only need to consider 
the edges in $G$ that are between different subsets in the collection $\{ H_{\it inv}^+(i) \mid i \in [r]^-\}$.

To handle edges between two given vertex subsets $H_{\it inv}^+(i)$ and $H_{\it inv}^+(j)$ in the graph $G$, 
we can employ an $\ell_0$-sampler algorithm (see Section 2.4), which, by Proposition~\ref{lem:sampler}, can 
handle dynamic edge changes and edge samplings between the two vertex subsets, efficiently in terms of both 
space complexity and update time. This, however, does not take edge weights into consideration, which can 
certainly impact the weight of the constructed $k$-matching. To address this issue, instead of using a single 
$\ell_0$-sampler for a pair of values in $[r]^-$, for each edge weight value $w$, and for each pair $(i, j)$ of 
values in $[r]^-$, we employ an $\ell_0$-sampler ${\bf L}_{i,j,w}$ to handle the dynamic changes of weight-$w$ 
edges between the two vertex subsets $H_{\it inv}^+(i)$ and $H_{\it inv}^+(j)$. Now, for an edge $[u, v]$ 
in the maximum weighted $k$-matching $M_{\max}$, the associated $\ell_0$-sampler ${\bf L}_{i, j, wt(u, v)}$, 
where $u \in H_{\it inv}^+(i)$, $v \in H_{\it inv}^+(j)$, and $H_{\it inv}^+(i)$ and $H_{\it inv}^+(j)$ are both 
in the collection $H_{\it inv}^+(B)$, will sample an edge of weight $wt(u, v)$ between $H_{\it inv}^+(i)$ and 
$H_{\it inv}^+(j)$. Since the subsets in the collection $H_{\it inv}^+(B)$ are pairwise disjoint, the $k$ 
$\ell_0$-samplers associated with the $k$ edges in $M_{\max}$ give a maximum weighted $k$-matching in $G$. 
Thus, the following collection of $\ell_0$-samplers 
\[ \{ {\bf L}_{i,j,w} \mid i, j \in [r]^-, \mbox{$w$ is an edge weight value in $G$} \} \] 
makes a sketch that is a subgraph of $G$ containing a maximum weighted $k$-matching in $G$. 

\begin{figure}[htbp]
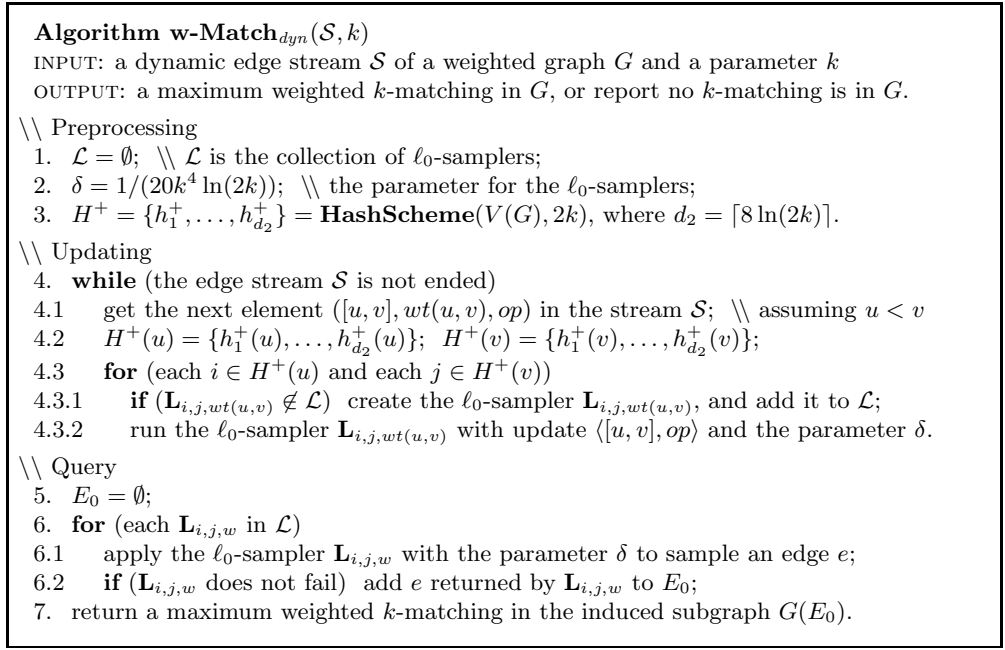

\setbox4=\vbox{\hsize30.5pc  \smallskip
 \noindent \strut 
\footnotesize 
\hspace*{2mm}{\bf  Algorithm {\bf w-Match$_{\it dyn}$}$(\SSS, k)$}\\
\hspace*{2mm}{\sc input}:  a dynamic edge stream $\SSS$ of a weighted graph $G$ and a parameter $k$\\
\hspace*{2mm}{\sc output}: a maximum weighted $k$-matching in $G$, or report no $k$-matching is in $G$. 

\smallskip

$\backslash\backslash$ Preprocessing\\
\hspace*{2mm}1. \  ${\cal L} = \emptyset$; \ $\backslash\!\backslash$ $\cal L$ is the collection of 
   $\ell_0$-samplers;\\
\hspace*{2mm}2. \  $\delta  = 1/(20k^4\ln (2k))$; \ $\backslash\!\backslash$ the parameter for 
    the $\ell_0$-samplers;\\
\hspace*{2mm}3. \  $H^+ = \{h_1^+, \ldots, h_{d_2}^+\} =  {\bf HashScheme}(V(G), 2k)$, where 
    $d_2 = \lceil 8 \ln (2k) \rceil$.

\smallskip

$\backslash\backslash$ Updating\\
\hspace*{2mm}4. \  {\bf while} (the edge stream $\SSS$ is not ended)\\
\hspace*{2mm}4.1 \hspace*{3mm} get the next element $([u, v], wt(u, v), op)$ in the stream $\SSS$; 
     \ $\backslash\!\backslash$ assuming $u < v$\\
\hspace*{2mm}4.2 \hspace*{3mm}  $H^+(u) = \{h_1^+(u), \ldots, h_{d_2}^+(u)\}$; \  
                           $H^+(v) = \{h_1^+(v), \ldots, h_{d_2}^+(v)\}$;\\ 
\hspace*{2mm}4.3 \hspace*{3mm}   {\bf for} (each $i \in H^+(u)$ and each $j \in H^+(v)$)\\
\hspace*{2mm}4.3.1 \hspace*{4mm} {\bf if} $({\bf L}_{i,j,wt(u,v)} \not\in {\cal L})$ \ 
     create the $\ell_0$-sampler ${\bf L}_{i,j,wt(u,v)}$, and add it to $\cal L$;\\ 
\hspace*{2mm}4.3.2 \hspace*{4mm}  run the $\ell_0$-sampler ${\bf L}_{i,j,wt(u,v)}$ 
    with update $\langle [u, v], op \rangle$ and the parameter $\delta$. 

\smallskip

$\backslash\backslash$ Query\\
\hspace*{2mm}5. \  $E_0 = \emptyset$;\\
\hspace*{2mm}6. \  {\bf for} (each ${\bf L}_{i,j,w}$ in $\cal L$)\\ 
\hspace*{2mm}6.1 \hspace*{3mm} apply the $\ell_0$-sampler ${\bf L}_{i,j,w}$ with the parameter 
    $\delta$ to sample an edge $e$;\\ 
\hspace*{2mm}6.2 \hspace*{3mm} {\bf if} (${\bf L}_{i,j,w}$ does not fail) \ 
     add $e$ returned by ${\bf L}_{i,j,w}$ to $E_0$; \\
\hspace*{2mm}7. \  return a maximum weighted $k$-matching in the induced subgraph $G(E_0)$. 
\medskip
\strut} $$\boxit{\box4}$$
 \vspace{-8mm}
\caption{A streaming algorithm for \wmatcheq{} in the dynamic model.} 
\label{chenfig4}
\end{figure}

The formal description of our algorithm is given in Figure~\ref{chenfig4}. Without loss of generality, 
we will assume that the algorithm is queried at the end of the stream $\SSS$, even though the query 
could take place at any point in the stream. 

\begin{lemma}  
\label{A_reduce_update}
If the graph $G$ contains $k$-matchings, then, with probability $\geq 1 - 11/(20k^3\ln (2k))$, the 
algorithm {\bf w-Match}$_{dyn} (\SSS, k)$ returns a  maximum weighted $k$-matching of the graph $G$.

\begin{proof}
Let  $M_{\max} = \{ [u_1, v_1], \ldots, [u_k, v_k] \}$ be a maximum weighted $k$-matching in the graph 
$G = (V, E)$, where $u_j < v_j$ for all $j$. From the algorithm {\bf HashScheme$(V(G), k)$}, as given in 
Figure~\ref{chenfig14}, for a vertex $u$ in the graph $G$, each value in the set $H^+(u)$ is in the range 
$[r]^-$, where $r = O(k \log^2 k) = O(k \cdot \plog)$. Recall that for each $i \in [r]^-$, $H_{\it inv}^+(i)$ is the set 
of vertices $u$ in $G$ such that $i \in H^+(u)$. By Theorem \ref{alg-hash-THM4}, with probability at least 
$1 - 4/((2k)^3 \ln (2k)) = 1 - 1/(2k^3 \ln (2k))$, there are $2k$ pairwise disjoint subsets $H_{\it inv}^+(i_1)$, 
$H_{\it inv}^+(i_1')$, $\ldots$, $H_{\it inv}^+(i_k)$, $H_{\it inv}^+(i_k')$ such that $u_j \in H_{\it inv}^+(i_j)$ 
and $v_j \in H_{\it inv}^+(i_j')$, for $1 \leq j \leq k$. Let $\mathcal{E}'$ be this event. Then 
$\Pr[\mathcal{E'}] \geq 1 - 1/(2k^3\ln (2k))$. Under the event $\mathcal{E}'$, for each $j$, $1 \leq j \leq k$, 
step~4.3.1 of the algorithm {\bf w-Match$_{\it dyn}$} will feed an edge of weight $wt(u_j, v_j)$ into the 
$\ell_0$-sampler ${\bf L}_{i_j, i_j', wt(u_j, v_j)}$. 

Now consider the sampling success probability for the $\ell_0$-samplers, under the condition of the event 
$\mathcal{E'}$. For each edge $[u_j, v_j]$ in the matching $M_{\max}$, we call the $\ell_0$-sampler 
${\bf L}_{i_j, i_j', wt(u_j, v_j)}$ in step 6.1 of the algorithm {\bf w-Match$_{\it dyn}$} to sample an edge between 
the two subsets $H_{\it inv}^+(i_j)$ and $H_{\it inv}^+(i_j')$. Let $\mathcal{E}''$ be the event that for all $j$, 
$1 \leq j \leq k$, an edge $e_j$ is sampled successfully by the $\ell_0$-sampler ${\bf L}_{i_j, i_j', wt(u_j, v_j)}$. 
Note that $e_j$ may not be the edge $[u_j, v_j]$ in the maximum weighted matching $M_{\max}$, but it must 
be an edge of weight $wt(u_j, v_j)$ between the two subsets $H_{\it inv}^+(i_j)$ and $H_{\it inv}^+(i_j')$. By 
Proposition~\ref{lem:sampler}, for each $j$, the $\ell_0$-sampler ${\bf L}_{i_j, i_j', wt(u_j, v_j)}$ fails with 
probability at most $\delta$, which gives, by the union bound, 
$\Pr[\mathcal{E}''] \geq 1 - k \cdot \delta  \geq 1 - 1/(20k^3\ln (2k))$, because $\delta = 1/(20k^4\ln (2k))$. 
Since the $2k$ subsets $H_{\it inv}^+(i_1)$, $H_{\it inv}^+(i_1')$, $\ldots$, $H_{\it inv}^+(i_k)$, 
$H_{\it inv}^+(i_k')$ are pairwise disjoint under the event $\mathcal{E'}$, the $k$ edges $e_1$, $\ldots$, $e_k$ 
sampled by the $k$ $\ell_0$-samplers ${\bf L}_{i_j, i_j', wt(u_j, v_j)}$, $1 \leq j \leq k$, respectively, share no 
common endpoints, i.e., the edge set $\{ e_1, \ldots, e_k\}$ is a $k$-matching in $G$. Moreover, because 
$wt(e_j) = wt(u_j, v_j)$ for all $1 \leq j \leq k$, $\{ e_1, \ldots, e_k\}$ is actually a maximum weighted 
$k$-matching in $G$. As a consequence, under the event $\mathcal{E}' \cap \mathcal{E}''$, the 
subgraph $G(E_0)$ of the graph $G$ induced by the edge set $E_0$ constructed in step 6.2 of the 
algorithm {\bf w-Match$_{\it dyn}$} contains a maximum weighted $k$-matching in the graph $G$. Finally, 
because $\Pr[\mathcal{E}'] \geq 1 - 1/(2k^3 \ln (2k))$ and $\Pr[\mathcal{E}''] \geq 1 - 1/(20k^3\ln (2k))$,  
by the union bound: 
\[ \Pr[\mathcal{E}' \cap \mathcal{E}''] \geq 1 - 1/(2k^3\ln (2k)) - 1/(20k^3\ln (2k)) = 1 - 11/(20k^3\ln (2k)). \]
In conclusion, under the event $\mathcal{E}' \cap \mathcal{E}''$, step 7 of the algorithm {\bf w-Match$_{\it dyn}$} 
returns a maximum weighted $k$-matching in the graph $G$. The lemma is proved. 
\end{proof}
\end{lemma}

Now we arrive at our conclusion for the algorithm {\bf w-Match$_{\it dyn}(\SSS, k)$}.

\begin{theorem}  
\label{theorem7}
Let $(\SSS. k)$ be a stream of a weighted graph $G$ in the dynamic model. Then
\vspace*{-2mm}
\begin{enumerate}
  \item[(1)] if $G$ contains a $k$-matching then, with probability at least $1 - 11/(20k^3\ln (2k))$, the 
    algorithm {\bf w-Match$_{\it dyn}(\SSS, k)$} returns a maximum weighted $k$-matching in $G$; and
\vspace*{-2mm}
  \item[(2)] if $G$ does not contain a $k$-matching then the algorithm {\bf w-Match$_{\it dyn}(\SSS, k)$} 
    reports so.
\end{enumerate}
\vspace*{-2mm}
Moreover, the algorithm {\bf w-Match$_{\it dyn}(\SSS, k)$} uses $O(k^2 W \cdot \plog)$ space and has 
$O(\plog)$ update time, where $W$ is the number of distinct edge weight values in the graph $G$.

\begin{proof}
First observe that the graph $G(E_0)$ in step 7 induced by the edge set $E_0$ constructed by step 6.2 of 
the algorithm {\bf w-Match$_{\it dyn}$} is a subgraph of $G$. Therefore, statement (2) in the theorem 
clearly holds true. Statement (1) follows from Lemma~\ref{A_reduce_update}. 

To analyze the complexities of the algorithm {\bf w-Match$_{\it dyn}(\SSS, k)$}, first recall that for a vertex 
$v$ in the graph $G$, the set $H^+(v)$ is a subset of $[r]^-$, where $r = d_1 d_2 d_3$, $d_1 = O(k/\ln k)$, 
$d_2 = O(\ln k)$, and $d_3 = O(\ln^2 k)$ (see the algorithm {\bf HashScheme} in Figure~\ref{chenfig14}). 
Thus, $r = O(k \ln^2 k)$. 

Consider the update time of the algorithm. The update on elements in the stream $\SSS$ is processed 
by steps 4.2-4.3. By Theorem~\ref{alg-hash-space}, step 4.2 takes time polynomial in $\log |V|$ 
(note $d_2 = O(\ln k)$). For step 4.3, since each of the subsets $H^+(u)$ and $H^+(v)$ contains 
$d_2 = O(\ln k)$ values, step 4.3 examines $O(\ln^2 k)$ pairs of the form $(i, j)$. We can organize 
all the $\ell_0$-samplers ${\bf L}_{i,j,w}$ in a 2-dimensional array $C[1..r, 1..r]$, where the element 
$C[i,j]$ is a balanced search tree for the weights of the edges between the subsets $H_{\it inv}^+(i)$ 
and $H_{\it inv}^+(j)$, which supports searching, insertion, and deletion in logarithmic time per 
operation \cite{Cormen}. The array  $C[1..r, 1..r]$ of space $O(r^2 W)$ supports searching 
a given $\ell_0$-sampler ${\bf L}_{i,j,w}$ in step 4.3.1 in time $O(\log W)$. This, plus the time for 
updating the $\ell_0$-sampler ${\bf L}_{i,j,w}$ in steps 4.3.1-4.3.2 (which is $O(\plog)$ by 
Proposition~\ref{lem:sampler}), shows that steps 4.3.1-4.3.2 take time $O(\plog)$. As a result, step 4.3 
takes time $d_2^2 \cdot O(\plog) = O(\plog)$ because $d_2 = O(\ln k)$. In conclusion, the update 
time of the algorithm {\bf w-Match$_{\it dyn}(\SSS, k)$} on each element in the stream $\SSS$, as given 
in steps 4.1-4.3 of the algorithm, is $O(\plog)$. 

We analyze the space complexity of the algorithm. By Theorem~\ref{alg-hash-space}, the space taken 
by steps 1-3 of the algorithm is $O(\log k \log |V|)$, which is used to initialize certain values and to 
construct and store the hash functions in $H^+$. For step 4, as we described above, we can use a 
2-dimensional array $C[1..r, 1..r]$ to store the $r^2 W$ $\ell_0$-samplers ${\bf L}_{i,j,w}$. Moreover, 
since $\delta = 1/(20k^4\ln (2k))$, by Proposition~\ref{lem:sampler}, each $\ell_0$-sampler ${\bf L}_{i,j,w}$ 
uses $O(\log^2 |V| \cdot \log k)$ space. As a result, step 4 of the algorithm totally takes space 
$O(r^2 W \log^2 |V| \log k)$. Now consider steps 5-7 of the algorithm. The space complexity of steps 5-7 
is dominated by the space used to store the $r^2 W$ $\ell_0$-samplers ${\bf L}_{i,j,w}$, which is bounded 
by $O(r^2 W \log^2 |V| \log k)$ as analyzed above, and the space used to store the induced subgraph 
$G(E_0)$. Since at most one edge is sampled from each $\ell_0$-sampler and there are $r^2 W$ 
$\ell_0$-samplers, the number of edges in the set $E_0$, thus the size of the graph $G(E_0)$, is 
$O(r^2 W)$. Finally, step 7 of the algorithm uses space $O(|E_0|) = O(r^2 W)$ to construct 
a  maximum weighted $k$-matching in the induced subgraph $G(E_0)$ \cite{gabow2}. Summarizing 
all the discussions above, we conclude that the space complexity of the algorithm {\bf w-Match$_{\it dyn}$} is 
$O(r^2 W \log^2 |V| \log k) = O(k^2 W \cdot \plog)$, thus, complete the proof of the theorem.
\end{proof}
\end{theorem}

{\bf Remark.} When $W = 1$, Theorem~\ref{theorem7} gives a streaming algorithm in the dynamic model 
for the \umatch{} problem, i.e., the $k$-matching problem on unweighted graphs. The algorithm runs in 
$O(k^2 \cdot \plog)$ space and $O(\plog)$ update time, and has success probability at least 
$1 - 11/(20k^3\ln (2k))$. The known lower bounds for \umatch{}, as we will discuss in the next section, 
show that both the space complexity and update time of this algorithm are nearly optimal, i.e., differing 
from the corresponding optimal bounds by at most a poly-logarithmic factor.

The space complexity $O(k^2 W \cdot \plog)$ of the streaming algorithm {\bf w-Match$_{\it dyn}$} in the 
dynamic model for weighted graphs is large if the number $W$ of distinct edge weights is large. 
Unfortunately, as we will prove in the appendix, the term $W$ in space complexity for streaming 
algorithms in the dynamic model for the \wmatcheq{} problem is actually unavoidable. On the other 
hand, as suggested in~\cite{Chitnis2016}, by rounding the edge weights, algorithms such as   
{\bf w-Match$_{\it dyn}$} can be used to develop streaming approximation algorithms in the dynamic 
model for the \wmatcheq{} problem. 

Under the Size-$k$ Constraint, an approximation streaming algorithm for the \wmatcheq{} problem was 
presented in~\cite{Chitnis2016}. The algorithm approximates \wmatcheq{} to within ratio $1 + \epsilon$, 
for any $\epsilon > 0$, and has space complexity $O(k^2 \log R_{\it wt} \cdot \plog/\epsilon)$ and update 
time $O(\plog)$, where $R_{\it wt}$ is the ratio of the maximum edge weight over the minimum edge weight. 
Approximation streaming algorithms for \wmatcheq{} in the dynamic model with no assumption of Size-$k$ 
Constraint were also studied and developed in \cite{Chitnis2016}, which are able to keep the same space 
complexity but have to worsen the update time to $O(k^2 \cdot \plog)$. 
 
Using Theorem~\ref{theorem7}, and following the same approach in~\cite{Chitnis2016}, we obtain 
the following approximation streaming algorithm of ratio $1+\epsilon$ for the \wmatcheq{} problem 
in the dynamic model. The algorithm has space complexity $O(k^2\log R_{\it wt} \cdot \plog/\epsilon)$ 
and update time $O(\plog)$, and does not need to assume the Size-$k$ Constraint.

\begin{theorem}  
\label{theorem8}
For any $0< \epsilon <1$, there is an algorithm for the \wmatcheq{} problem in the dynamic model that 
on a stream $(\SSS, k)$ for a weighted graph $G$:
\vspace*{-1mm}
\begin{enumerate}
  \item[(1)]  returns a $k$-matching of weight at least $(1 - \epsilon)$ of that of a maximum weighted 
   $k$-matching in $G$ if $G$ contains $k$-matchings; and 
\vspace*{-1mm}
  \item[(2)] reports no $k$-matching if $G$ does not contain a $k$-matching.
\end{enumerate}
\vspace*{-1mm}
Moreover, the algorithm uses $O(k^2 \log R_{\it wt} \cdot \plog / \epsilon)$ space and has $O(\plog)$ 
update time, where $R_{\it wt}$ is the ratio of the maximum edge weight over the minimum edge weight.

\begin{proof}
For each edge $e$ in the graph $G$, we assign $e$ a new weight $wt'(e) = t$, where $t$ is the 
integer satisfying $(1+\epsilon)^{t-1} < wt(e) \leq (1+\epsilon)^t$. Thus, under the new edge weights, 
the graph $G$ has $O(\log R_{\it wt} / \epsilon)$ distinct edge weights. Now we run the algorithm 
{\bf w-Match$_{\it dyn}$} on the graph $G$ with the new edge weights. By Theorem~\ref{theorem7}, 
the algorithm returns a $k$-matching $M$ in $G$ in space $O(k^2\log R_{\it wt} \cdot \plog/ \epsilon)$ and 
update time $O(\plog)$, with success probability at least $1 - 11/(20k^3\ln(2k))$.

We prove that in terms of the original edge weight function $wt(\cdot)$ of the graph $G$, the weight 
$wt(M)$ of the $k$-matching $M$ returned by the above algorithm is at least $(1 - \epsilon)$ of the 
weight $wt(M_{\max})$ of a maximum weighted $k$-matching $M_{\max} = \{e_1, \ldots, e_k\}$ in the 
graph $G$. Consider the algorithm {\bf w-Match$_{\it dyn}$} on the graph $G$ with the new edge weights. 
As proved in Lemma~\ref{A_reduce_update}, with probability at least $1 - 11/(20k^3\ln (2k))$, the induced 
subgraph $G(E_0)$ constructed by step 6 of the algorithm {\bf w-Match$_{\it dyn}$} contains a 
$k$-matching $M' = \{ e_1', \ldots, e_k' \}$, where for each $s$, the edge $e_s'$ in $M'$ and the edge 
$e_s$ in the maximum weighted $k$-matching $M_{\max}$ are processed by the same $\ell_0$-sampler 
${\bf L}_{i,j,w}$. Therefore, under the new edge weights of the graph $G$, the edges $e_s'$ and $e_s$ 
have the same weight $wt'(e_s') = wt'(e_s) = w$, which, by the definition of the new edge weights, 
immediately gives the relation $wt(e_s)/wt(e_s') \leq 1 + \epsilon$ on the original edge weights of the 
graph $G$. Summarizing this over all $1 \leq s \leq k$, we get 
\[ \frac{wt(M_{\max})}{wt(M')} = \frac{\sum_{s=1}^k wt(e_s)}{\sum_{s=1}^k wt(e_s')}  \leq 1 + \epsilon. \]
Therefore, with probability $\geq 1 - 11/(20k^3\ln(2k))$, the induced subgraph $G(E_0)$ constructed by 
step 6 of the algorithm {\bf w-Match$_{\it dyn}$} contains the $k$-matching $M'$ whose weight $wt(M')$ 
is at least $wt(M_{\max})/(1 + \epsilon)$. As a result, the $k$-matching $M$ returned by the algorithm 
{\bf w-Match$_{\it dyn}$}, which is a maximum weighted $k$-matching in $G(E_0)$ in terms of the new 
edge weights, has its weight $wt(M)$ at least $wt(M_{\max})/(1 + \epsilon) \geq (1 - \epsilon) wt(M_{\max})$. 
This completes the proof of the theorem.
\end{proof}
\end{theorem}

\section{Conclusions and final remarks}
\label{senclusion}

In this paper, we presented streaming algorithms for the fundamental graph $k$-matching problem, for 
both unweighted graphs and weighted graphs, and in both the insert-only and dynamic streaming models. 
While matching the best space complexity of known algorithms, our algorithms have much faster update 
times, significantly improving previous known results. In fact, our algorithms are optimal or near-optimal 
for many cases for the graph $k$-matching problem. We give below a brief discussion on the optimality 
of our algorithms when they are applied in various cases of the graph $k$-matching problem. 

A lower bound $\Omega(k^2)$ on the space complexity for the \umatch{} problem in the insert-only 
streaming model to construct a $k$-matching in an unweighted graph has been developed in 
\cite{Chitnis2016}, which shows that for any randomized streaming algorithm for the problem, there are 
instances of size $n$ and parameter $k$, such that the algorithm takes space of $\Omega(k^2)$ bits. 
The more difficult problem \wmatcheq{} in the insert-only streaming model is to construct a maximum 
weighted $k$-matching in a weighted graph, for which the $\Omega(k^2)$ space lower bound certainly 
holds true. By Theorem~\ref{thm:insert}, our streaming algorithm {\bf w-Match$_{\it ins}$} given in section 
3 solves the \wmatcheq{} problem in the insert-only model in space $O(k^2)$ and update time $O(1)$. 
The optimality of the update time $O(1)$ of the algorithm {\bf w-Match$_{\it ins}$} is obvious. Thus, the 
streaming algorithm {\bf w-Match$_{\it ins}$} solves the \wmatcheq{} problem in the insert-only model in 
both optimal space complexity and optimal update time. To the authors' best knowledge, this is the first 
streaming algorithm for the \wmatcheq{} problem that achieves optimality in both space complexity and 
update time complexity.

Similarly, the \umatch{} problem in the dynamic streaming model is at least as hard as the problem in the 
insert-only streaming model, so the space lower bound $\Omega(k^2)$ also holds true for the \umatch{} 
problem in the dynamic model. As we remarked in the paragraph following Theorem~\ref{theorem7}, our 
streaming algorithm {\bf w-Match$_{\it dyn}$} given in section 4 has space complexity $O(k^2 \cdot \plog)$ and 
update time $O(\plog)$ when it is applied in solving the \umatch{} problem in the dynamic model. This 
presents the first streaming algorithm in the dynamic model that solves the \umatch{} problem in both 
near-optimal space complexity and near-optimal update time complexity, where by ``near-optimal'', we mean 
that the complexity bounds of the algorithm differ from the corresponding optimal bounds only by a 
poly-logarithmic factor. 

By Theorem~\ref{theorem7}, when the algorithm {\bf w-Match$_{\it dyn}$} is applied to solve the 
\wmatcheq{} problem in the dynamic model, it still keeps the near-optimal update time $O(\plog)$, but 
increases its space complexity to $O(k^2 W \cdot \plog)$, which will be quite significant if the number 
$W$ of distinct edge weight values is large (note that $W$ can be as large as the number of edges in 
the input graph). Unfortunately, the dependency of the space complexity on the value $W$ for streaming 
algorithms solving the problem \wmatcheq{} is actually unavoidable: in the appendix, we give a proof 
that any randomized streaming algorithm in the dynamic model that solves the \wmatcheq{} problem 
has space complexity $\Omega(\max\{W \log W, k^2\})$. Readers who are interested in space lower 
bounds for streaming algorithms are referred to \cite{dark,kapralov} for more recent developments.

We believe that the hash scheme we developed in subsection 4.1 is of independent interests. Different 
from the standard hashing techniques that {\it partition} the universal set $U$ into pairwise disjoint 
subsets, our hash scheme is actually a many-to-many mapping from the universal set 
$U$ to a set whose size is smaller than that for standard hashing. Therefore, our hash scheme 
constructs a collection of (unnecessarily disjoint) subsets of $U$, but ensures that a subcollection of 
disjoint subsets distinguishes a subset of $k$ elements in $U$. Compared with the standard hashing 
techniques, this approach uses less space and has a higher success probability. We believe that the 
results and techniques can have wider applications in other fundamental graph problems. 

When applied to the problem \umatch{} for unweighted graphs, the space complexity and update time 
of our streaming algorithm {\bf w-Match$_{\it dyn}$} in the dynamic model are near-optimal, which still 
differ from the corresponding proven lower bounds by a poly-logarithmic factor. When the algorithm is 
applied to the problem \wmatcheq{} for weighted graphs in the dynamic model, the gap between the 
upper bound provided by the algorithm {\bf w-Match$_{\it dyn}$} and the proven lower bound is still 
quite significant. It will be interesting to study how much we can further narrow down or even 
close the gaps between the upper bounds and the lower bounds. In particular, is it possible to have 
streaming algorithms in the dynamic model for the \umatch{} problem with space complexity $O(k^2)$ 
and update time $O(1)$? This question is also related to the space lower bounds on streaming 
approximation algorithms for maximum matching in the dynamic model \cite{dark,konrad}.

Another interesting problem that deserves further study is the trade-off between the space complexity 
and the update time of streaming algorithms. The $O(1)$ update time of our streaming algorithm 
{\bf w-Match$_{\it ins}$} for the \wmatcheq{} problem in the insert-only model used a technique of 
interleaving the process of updating a sketch, which is the subgraph $G_s^f$ of the input graph $G$ 
in our algorithm, with the process of reading the next input stream segment $G_{s+1,s'}$ (see 
Figure~\ref{chenfig1}). To make the time for reading the new input stream segment to ``cover'' that for 
updating the sketch, smaller memory space for storing the (thus, shorter) new input stream segment 
would require longer update time per element in the segment, while faster update time per element 
in the stream would result in reading a longer new input stream segment that requires larger memory 
space for storing the segment. Moreover, longer new input stream segment would make the sketch 
updating more time consuming to include the information brought in by the longer new segment. It 
would be interesting to study the interaction/relation between these parameters in streaming algorithms. 

\bibliography{arXivRef}

\appendix

\section{Lower Bounds}
\label{subsec:lowerbounds}

In this appendix, we study lower bounds on the space complexity of randomized streaming algorithms 
for the problems \umatch{} and  \wmatcheq{}, in both the insert-only model and the dynamic model. 
These lower bound results, in conjunction with the algorithms given in this paper, show that in many 
cases, the space complexity achieved by our algorithms is optimal or near-optimal (i.e., optimal modulo 
a poly-logarithmic factor in the input size). Note that the update times of our algorithms, which are 
$O(1)$ in the insert-only model and $O(\plog)$ in the dynamic model, are already optimal or 
near-optimal. 

A lower bound $\Omega(k^2)$ on the space complexity for \umatch{} in the insert-only model has 
been developed in \cite{Chitnis2016}, which shows that for any randomized streaming algorithm for 
the problem, there are instances of size $n$ and parameter $k$, where $n = \Theta(k^2)$, such that the 
algorithm takes space of $\Omega(n) = \Omega(k^2)$ bits. This result does not seem to address the 
following issues that are special for parameterized streaming algorithms in which (1) the graph size 
$n$ and the parameter $k$ in general are relatively independent, and (2) the graph size $n$ can be 
much larger than the parameter $k$.

In the following, we introduce a new definition for lower bounds for streaming problems, which tries to 
address the above issues that are special for parameterized streaming algorithms. The definition is given 
in terms of space complexity, but can be extended to other complexity measures.

\begin{definition}
A parameterized streaming problem $Q$ has a {\it space complexity lower bound $\Omega(g(k))$} if 
for any streaming algorithm $A$ for $Q$, there are infinitely many parameter values $k$ such that for 
each such parameter value $k$ and for any integer $n \geq k$ ($n$ does not depend on $k$), there 
are instances of parameter $k$ and size larger than $n$ on which the algorithm $A$ runs in space 
$\Omega(g(k))$.
\end{definition}  

The definition above is consistent with the standard ones. In particular, a lower bound in terms of this  
definition implies the same lower bound in terms of the standard definition. 

As in the previous work such as \cite{Chitnis2016}, we will use the \emph{one-way communication 
model} to prove lower bounds on the space complexity of randomized streaming algorithms for 
\umatch{} and \wmatcheq. In this model, there are two parties, Alice and Bob, each receiving $x$ and 
$y$, respectively, who wish to compute $f(x,y)$. Alice is permitted to send Bob a single message $M$, 
which only depends on $x$ and Alice's random coins. Then Bob outputs $b$, which is his guess of 
$f(x,y)$. Here, $b$ only depends on $y,M,$ and Bob's random coins. We say the protocol computing 
$f$ with success probability $1-\delta$ if $\Pr[b = f(x,y)] \ge 1-\delta$ for every $x$ and $y$. 

\subsection{Lower bounds for \umatch}

We will use the lower bound on the communication complexity of the following problem:
\begin{quote} 
The {\sc Index} problem: Alice has an $m$-bit string $x \in \{0,1\}^m$ and Bob has an integer $z\in [m]$. 
The goal is to compute the $z$-th bit of $x$. 
\end{quote}
It is known \cite{kremer1999randomized} that any randomized communication protocol solving the 
{\sc Index} problem with success probability $\geq 2/3$ has communication complexity of $\Omega(m)$ 
bits. The constant $2/3$ can be improved to any constant strictly greater than $1/2$ \cite{tim}. 

\begin{theorem}
\label{thm:lowerboundinsert}
Any randomized streaming algorithm for the \umatch{} problem in the insert-only model with 
success probability at least $2/3$ uses $\Omega(k^2)$ bits of space.

\begin{proof}
The proof follows the ideas of \cite{Chitnis2016}, with modifications to meet the additional conditions 
in the lower bound definition given above. Let ${\bf A}_{\it match}$ be any randomized streaming 
algorithm for \umatch{} in the insert-only model with success probability at least $2/3$. We show how 
to use the algorithm ${\bf A}_{\it match}$ to construct a communication protocol for the {\sc Index} 
problem. Let $(x, z)$ be an instance of the {\sc Index} problem, where $x \in \{0,1\}^m$ and $z \in [m]$. 
Define a subset of $[m]$ as $S = \{ i \mid \mbox{the $i$-th bit of $x$ is $1$} \}$. The task is to
decide whether or not $z \in S$. 
 
Let $k_1 = \lceil \sqrt{m}\, \rceil$. Fix an injection $\chi: [m] \longrightarrow  [k_1] \times [k_1]$. 
Suppose $\chi(z) = (p_z, q_z)$. Construct a graph $G$ whose vertex set contains two disjoint subsets 
of $2k_1$ vertices: $V_L = \{l_i, l_i^* \mid i \in [k_1]\}$ and $V_R = \{r_i, r_i^* \mid i \in [k_1]\}$. 
The edge set of $G$ contains the following edge subsets:

\smallskip 

(1) $E_S = \{ [l_s^*, r_t^*] \mid \mbox{if $(s, t) = \chi(y)$ for some $y \in S$} \}$; and  

(2) $E_L = \{ [l_s, l_s^*] \mid s \neq p_z \}$, and $E_R = \{ [r_t, r_t^*] \mid t \neq q_z \}$.

\smallskip

\noindent To make $G$ a graph of at least $n$ vertices for any $n > 4k_1$, we add to the graph $G$ 
a disjoint star of $n - 4k_1$ vertices given by the edge set 
$E_{\it star} = \{ [v_0, v_i] \mid 1 \leq i \leq n-4k_1-1 \}$. This completes the structure of the graph $G$, 
which has $n$ vertices, where $n > 4k_1$ can be any integer. It is not difficult to verify that the graph 
$G$ has a $(2k_1)$-matching if and only if $z \in S$. Now construct an instance $(\SSS, k)$ for \umatch{}, 
where $k = 2k_1$ and $\SSS$ is a stream for the graph $G$ that first inserts the edges in the set 
$E_S \cup E_{\it star}$, then inserts the edges in the set $E_L \cup E_R$. 

A communication protocol for the instance $(x, z)$ for the {\sc Index} problem works as follows: (1) Alice 
runs the streaming algorithm ${\bf A}_{\it match}$ for \umatch{} on the instance $(\SSS, k)$ for the 
elements in $\SSS$ that insert the edges in $E_S \cup E_{\it star}$ (Alice can generate these elements 
because she knows $S$). After processing all edges in $E_S \cup E_{\it star}$, Alice sends the memory 
contents of her computation to Bob. Bob then uses the memory contents of Alice's computation and 
continues running the streaming algorithm ${\bf A}_{\it match}$ on the rest of the elements in the stream 
$\SSS$ that insert the edges in $E_L \cup E_R$ (Bob can generate these elements because he knows 
$z$). As we observed, Bob claims $z \in S$ if and only if the algorithm ${\bf A}_{\it match}$ on $(\SSS, k)$ 
returns a $k$-matching in $G$. Thus, by the assumptions on the algorithm ${\bf A}_{\it match}$, this is a 
randomized communication protocol solving the {\sc Index} problem with probability $\geq 2/3$. By 
\cite{kremer1999randomized}, the communication complexity of the protocol is of $\Omega(m)$ bits.
Since the communication complexity of the protocol is equal to the size of the message Alice passed 
to Bob, which is the space used by Alice when she runs the algorithm ${\bf A}_{\it match}$, we conclude 
that the algorithm ${\bf A}_{\it match}$ on the instance $(\SSS, k)$ uses space of $\Omega(m) = \Omega(k^2)$ 
bits. The theorem is completed because ${\bf A}_{\it match}$ is an arbitrary algorithm for 
\umatch{}, and the size of the stream $\SSS$ can be arbitrarily larger than the parameter $k$. 
\end{proof}
\end{theorem}

Obviously, the space lower bound given by Theorem~\ref{thm:lowerboundinsert} also applies to 
streaming algorithms for the \umatch{} problem in the dynamic model.

\subsection{Lower bounds for \wmatcheq{}}

We start with the necessary background in information theory. For more details, see \cite{IT-book}.

For a random variable $X$, the (\emph{Shannon}) \emph{entropy} $H(X)$ of $X$ is defined as  
\[ H(X) = - \sum_x \Pr[X=x] \cdot \log (\Pr[X=x]). \] 
The \emph{binary entropy function} $H(q)$ is $H(X)$ for a 0-1 random variable $X$ with 
$\Pr[X = 1] = q$.

For two random variables $Z_1$ and $Z$, the \emph{conditional entropy $H(Z_1 \mid Z)$ of $Z_1$ 
given $Z$} is 
\[ H(Z_1 \mid Z) = \sum_z H(Z_1 \mid Z=z) \cdot \Pr[Z=z], \] 
and the \emph{mutual information} $I(Z_1; Z)$ is 
\[ I(Z_1; Z) = H(Z) - H(Z \mid Z_1). \] 
For random variables $Z_1$, $Z_2$, and $Z$, the \emph{conditional mutual information 
$I(Z_1; Z_2 \mid Z)$ of $Z_1, Z_2$ given $Z$} is 
\[ I(Z_1; Z_2 \mid Z) = H(Z_1 \mid Z) - H(Z_1 \mid Z_2, Z). \]

Random variables $X$, $Y$, $Z$ form a \emph{Markov chain} in that order (denoted 
by $X \rightarrow Y \rightarrow Z$) if the conditional distribution of $Z$ depends only on $Y$ and is 
conditionally independent of $X$. 

\begin{proposition}[Theorem 2.6.4, \cite{IT-book}]
\label{fact2}
For any random variable $X$, $H(X) \leq \log |\mathcal{X}|$, where $\mathcal{X}$ is the range of $X$, 
with equality if and only if $X$ has a uniform distribution over $\mathcal{X}$. 
\end{proposition}

\begin{proposition}[Theorem 2.5.2, \cite{IT-book}] 
\label{chain-rule}
For random variables $Z_1$, $Z_2$, $\ldots$, $Z_n$, and $Z$, we have the following chain rule: 
$I(Z_1, Z_2, \ldots, Z_n; Z) = \sum_{i=1}^n I(Z_i; Z \mid Z_{i-1}, Z_{i-2}, \ldots, Z_1)$. 
\end{proposition}

\begin{proposition}[Fano's Inequality, \cite{IT-book}, p.~39] 
\label{Fano}
Given a Markov chain $X \rightarrow Y \rightarrow X'$, and let $p = \Pr[X \neq X']$, then
$H(X \mid Y) \leq H(p) + p \cdot \log (|\mathcal{X}|-1)$, where $\mathcal{X}$ is the range of $X$.
\end{proposition}

Now we are ready for the lower bound for \wmatcheq{}. Consider the following problem:  
\begin{quote}
{\sc Partial Maximization}: Alice has a sequence $A = \langle a_1, a_2, \ldots, a_m \rangle$ of 
numbers in $[m^2]$, and Bob is given a set $P_B = \{(i, a_i) \mid i \in B\}$ of pairs, where $B$ 
is a subset of $[m]$. The goal is to compute $\max\{ a_t \mid t \not\in B \}$, i.e., to compute the 
largest number $a_t$ in the sequence $A$ that is not given to Bob. 
\end{quote}

\begin{theorem}
\label{llooww}
For any constant $\delta$, $0 \leq \delta <1$, any randomized one-way communication protocol for 
the {\sc Partial Maximization} problem with success probability at least $1-\delta$ has communication 
complexity of $\Omega(m \log m)$ bits, where $m$ is the length of the sequence given to Alice. The 
lower bound holds even when we assume that all numbers in the sequence given to Alice are distinct.

\begin{proof}
The proof is similar to that for the {\sc Augmented Indexing} problem \cite{augm1,augm2}. 
For each $j \in [m]$, consider the random variable $X_j$ that picks its value uniformly at random from 
$\{(j-1)m+1, \ldots, j \cdot m \}$. Then $X_1 < X_2 < \cdots < X_m$ - so all numbers in the sequence given 
to Alice are distinct, and $H(X_j) = \log m$ for all $j$. For each $j \in [m]$, let $B_j = \{ j+1, \ldots, m \}$ 
and let $X_j'$ be Bob's guess of $\max_{i \not\in B_j} X_i = X_j$. Let $M$ be the message sent from 
Alice to Bob. Since $\Pr[X'_j = X_j] \geq 1-\delta$ and 
$X_j \rightarrow (M,X_{j+1}, X_{j+2}, \ldots, X_m) \rightarrow X_j'$ is a Markov chain, by 
Proposition~\ref{Fano}, for all $j \in [m]$, we have 
\[ H(X_j \mid M, X_{j+1}, \ldots, X_m) \leq \delta \cdot \log m + 1. \]
Hence, 
\begin{eqnarray*}  
I(X_j; M \mid X_{j+1}, \ldots, X_m) 
      & = &    H(X_j \mid X_{j+1}, \ldots, X_m) - H(X_j \mid M,X_{j+1}, \ldots, X_m)\\
      &=&      H(X_j) - H(X_j \mid M,X_{j+1}, \ldots, X_m) \\
      &\geq& (1-\delta) \log m - 1,  
\end{eqnarray*}
where the second equality holds because $X_j$, $X_{j+1}$, $\ldots$, $X_m$ are mutually independent. 
By the definition of the mutual information and using the chain rule (Proposition~\ref{chain-rule}),
\begin{eqnarray*}
H(M) &\geq& I(X_1, X_2, \ldots, X_m; M)  \\   
 &=& \sum_{j=1}^m I(X_j; M \mid X_{j+1}, \ldots, X_m) \\ 
 &=&\Omega(m \log m).
\end{eqnarray*}
Finally, by Proposition~\ref{fact2}, $\log |{\cal M}| \geq H(M) = \Omega(m \log m)$, where $\cal M$ is the 
range of the message $M$, i.e., the message $M$ has at least $2^{\Omega(m \log m)}$ possibilities. As 
a result, the length of the longest possible message $M$ sent from Alice to Bob is $\Omega(m \log m)$.
\end{proof}
\end{theorem}

A space lower bound for streaming algorithms of the \wmatcheq{} problem in the dynamic model 
now can be derived by reducing {\sc Partial Maximization} to \wmatcheq{}. 

\begin{theorem}
\label{thm:lowerboundwmatching}
Let $0 \leq \delta<1$ be any constant. Any randomized streaming algorithm in the dynamic model for 
the \wmatcheq{} 
problem that, with probability $\geq 1-\delta$, computes a maximum weighted 1-matching has space 
complexity of $\Omega(W \log{W})$ bits, where $W$ is the number of distinct edge weights in the 
graph stream.

\begin{proof}
Let ${\bf A}_{\it match}$ be any randomized streaming algorithm for \wmatcheq{} that, with probability 
$\geq 1-\delta$, computes a maximum weighted 1-matching. We show how to use ${\bf A}_{\it match}$ 
to develop a communication protocol for {\sc Partial Maximization}. Let $(A, P_B)$ be an instance of 
{\sc Partial Maximization}, where $A = \langle a_1, a_2, \ldots, a_m \rangle$ is the sequence given to 
Alice, in which all numbers are distinct, and $P_B = \{(i, a_i) \mid i \in B\}$ is the set of pairs given to 
Bob, $B \subseteq [m]$.

Let $\SSS$ be a dynamic graph stream that first inserts $m$ arbitrary but distinct edges 
$\{ e_1, \ldots, e_m \}$, where for each $i$, the edge $e_i$ has weight $a_i$, then deletes the edges 
$e_i$ for $i \in B$. Obviously, a maximum weighted $1$-matching in the graph $G$ of the stream 
$\SSS$ has its weight equal to $\max\{ a_t \mid t \not\in B \}$, which is the solution to the instance 
$(A, P_B)$ of {\sc Partial Maximization}.

The communication protocol for {\sc Partial Maximization} on the instance $(A, P_B)$ works as 
follows: (1) Alice runs the streaming algorithm ${\bf A}_{\it match}$ on the instance $(\SSS, 1)$ of 
\wmatcheq{} for the first $m$ elements of edge insertions in the stream $\SSS$ (Alice can generate 
these elements because she knows the sequence $A$), then sends the memory contents of her 
computation to Bob; (2) After receiving the message from Alice, Bob uses the memory contents from 
Alice and continues running the algorithm ${\bf A}_{\it match}$ on $(\SSS, 1)$ for the rest elements in 
the stream $\SSS$, which are edge deletions (Bob can generate these elements because he knows 
the subset $B$), to get the maximum weighted $1$-matching, thus the solution to the instance 
$(A, P_B)$ of the {\sc Partial Maximization} problem. Under the assumption of the algorithm 
${\bf A}_{\it match}$, this gives a randomized communication protocol with success probability 
$\geq 1 -\delta$ for {\sc Partial Maximization}. By Theorem~\ref{llooww}, the size of the message sent 
from Alice to Bob, which is not larger than the space complexity of the algorithm ${\bf A}_{\it match}$ 
on $(\SSS, 1)$, has $\Omega(m \log m)$ bits. Moreover, by our assumption on the sequence $A$, 
$m$ is equal to the number $W$ of distinct edge weights in the stream $\SSS$. As a conclusion, the 
algorithm ${\bf A}_{\it match}$ on the instance $(\SSS, 1)$ uses space of $\Omega(W \log W)$ bits. 

Note that here the number $W$ of distinct edge weights is the parameter for which we have derived 
a lower bound. To make the lower bound hold true for graphs of size larger than $W$, i.e., to meet 
the additional conditions in the lower bound definition given at beginning of this section, we can simply 
let Alice add (many) more elements in the stream $\SSS$ that insert edges of weight equal to the 
smallest value in the sequence $A$.  
\end{proof}
\end{theorem}

Since the space lower bound $\Omega(k^2)$ for streaming algorithms for the \umatch{} problem in the 
insert-only model certainly applies for the \wmatcheq{} problem in the dynamic model, we obtain the 
following corollary. 

\begin{corollary}
Any randomized streaming algorithm for the \wmatcheq{} problem with success probability $\geq 2/3$ 
in the dynamic model uses $\Omega(\max\{ W \log W, k^2 \})$ bits of space, where $W$ is the number of 
distinct edge weights in the graph stream. 
\end{corollary}

\end{document}